\documentclass[11pt]{article}
\usepackage{geometry}
\usepackage{graphicx}
\usepackage{amssymb, amsmath}
\usepackage{epstopdf}
\usepackage{subfigure}
\usepackage{color}
\usepackage[table]{xcolor}
\usepackage{tikz}

\usetikzlibrary{backgrounds}

\definecolor{light}{RGB}{199, 153, 199}
\definecolor{dark}{RGB}{143, 39, 143}
\definecolor{gray80}{gray}{0.8}

\graphicspath{{figures/}}

\title{A Unified Treatment of Predictive Model Comparison}

\author{Michael Betancourt \\ Department of Statistics, University of Warwick}

\begin{document}
\maketitle

\begin{abstract}
The predictive performance of any inferential model is critical to
its practical success, but quantifying predictive performance is
a subtle statistical problem.  In this paper I show how the natural
structure of any inferential problem defines a canonical measure
of relative predictive performance and then demonstrate how
approximations of this measure yield many of the model comparison
techniques popular in statistics and machine learning.
\end{abstract}

Because any inferential method is built upon assumptions, one of the most 
important aspects of any statistical analysis is assessing the validity of the
underlying assumptions.  Although there are few model assessment approaches
that claim to validate a model in isolation, there is a rich history of 
\textit{comparative} techniques in the statistics literature,  from visual residual 
analyses to scoring rules and predictive cross validation to the myriad of 
information criteria.  The practical challenge in applying these methods, 
however, is in determining the ultimate accuracy of their assessments and 
hence which might be most appropriate to a given problem.

In this paper I demonstrate that any inferential system admits a canonical
measure of comparative predictive performance, here termed 
a \textit{relative predictive performance score}.  Moreover, I show how many 
of the model comparison techniques in practice today arise as approximations
to these canonical scores.  This foundational perspective provides a common
context for understanding the advantages and disadvantages of each technique
both in theory and in practice.

After reviewing the basic assumptions common to most inferential techniques, 
in particular the assumptions of frequentist and Bayesian inference, I demonstrate 
first how relative predictive performance scores arise naturally from these 
assumptions and then how various approximations of these scores yield existing 
model comparison techniques.  In the latter I emphasize how this foundational 
construction immediately identifies the practical consequences of each 
approximation.

\section{Foundational Assumptions of Inference}

The most fundamental assumption underlying any attempt at
inference is the existence of some \text{latent data generating process},
$\pi$, responsible for generating measurements.  Formally we
assume that measurements are drawn from some measurable space,
$\left( Y, \mathcal{Y} \right)$, and that the data generating process 
itself can be modeled mathematically by a single, time-invariant 
probability measure,
\begin{equation*}
\pi : \mathcal{Y} \rightarrow [0, 1].
\end{equation*}
In order to make this construction as general as possible I impose
no additional structure, such as a particular topology or metric, on 
$Y$, nor make any philosophical interpretation of this latent data 
generating process, in particular its interpretation as an ontological 
truth or just an epistemological impression.  Consequently the following 
results will hold regardless of any deeper meaning of the measurement 
process itself.

Inference is then a formal effort to learn the latent data generating 
process given a measurement, $y \in Y$, by identifying $\pi$ from 
the space of all probability measures, $P$, on the measurement space.  
Unfortunately, exploring the entirety of $P$ for any problem is far too 
unwieldly, and in order to construct practical inferential methods we first 
have to limit ourselves to a more manageable space of data generating 
processes.

An \textit{inferential model} is the selection of a distinguished subset of
data generating processes, $X \subset P$; in the spirit of Dennis
Lindley I refer to such subsets as \textit{small worlds} \cite{Lindley:2014}.  
As with the latent data generating process, I am careful not to assign any 
particular meaning to the small world -- it can be a phenomenological 
model motivated by mathematical and practical convenience, a theoretical 
model motivated by a specific scientific hypothesis, or, as is most common 
in practice, a delicate combination of the two.  In this paper I denote the 
measure corresponding to a given element of the small world, $x \in X$, as
\begin{equation*}
\pi_{x} : \mathcal{Y} \rightarrow [0, 1].
\end{equation*}

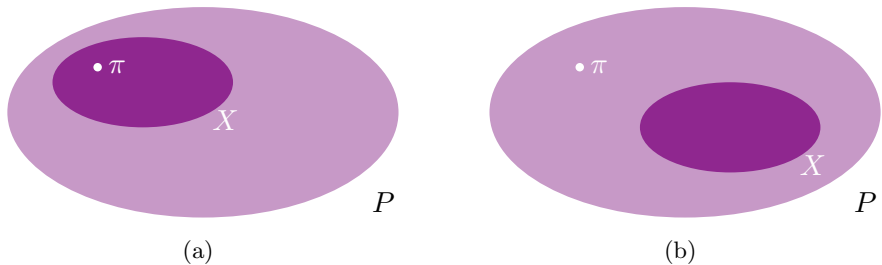
\begin{figure*}
\centering
\subfigure[]{
\begin{tikzpicture}[scale=0.20, thick]
  
  \draw[color=white] (-15, 0) -- (15, 0);
  
  \fill[light] (0, 0) ellipse (13 and 7);
  \node at (12, -6) {$P$};
  
  \fill[dark] (-4, 2) ellipse (6 and 3);
  \node[color=white] at (1.5, -0.5) {$X$};
  
  \fill[color=white] (-7, 3) circle (8pt)
  node[right, color=white] {$\pi$};
  
\end{tikzpicture}
}
\subfigure[]{
\begin{tikzpicture}[scale=0.20, thick]

  \draw[color=white] (-15, 0) -- (15, 0);

  \fill[light] (0, 0) ellipse (13 and 7);
  \node at (12, -6) {$P$};
  
  \fill[dark] (3, -1) ellipse (6 and 3);
  \node[color=white] at (8.5, -3.5) {$X$};
  
  \fill[color=white] (-7, 3) circle (8pt)
  node[right, color=white] {$\pi$};
  
\end{tikzpicture}
}
\caption{ Inference requires the selection of a distinguished subset of data 
generating processes that (a) may or (b) may not contain the latent data 
generating process, $\pi$.  The Boxian philosophy asserts that the former 
is impossible in practical problems but that we may still hope that some data 
generating process in $X$ will well-approximate $\pi$.
}
\label{fig:small_worlds}
\end{figure*}

One assumption that I have explicitly \textit{not} made is that the chosen 
small world need contain the latent data generating process 
(Figure \ref{fig:small_worlds}). In particular, assuming that the small 
world rarely, if ever, contains the latent data generation process formalizes 
the Boxian philosophy that ``all models are wrong but some are useful'' 
\cite{BoxEtAl:1987}.  Although developments in computation and theory,
such as statistical nonparametrics, have enabled the construction of 
increasingly complex models and less-small worlds, the intricacy of
any realistic measurement should continue to inspire skepticism in the
sufficiency of any small world.

From this perspective, the ultimate utility of any inferential procedure is
not in whether it can find the latent data generating process exactly
but rather in how well it can approximate the latent data generating process.
In particular, the fidelity of a procedure is often judged based on its
\textit{predictive performance}.  Exactly how we define predictive 
performance depends intimately on how inference is implemented, 
which itself depends on the fundamental interpretation of probability.

\section{Inference in the Small World}

We have already assumed that probability theory adequately describes the 
measurement process, but that does not have to be the only application of 
probability theory.  While \textit{frequentist inference} limits probabilities to 
the data, \textit{Bayesian inference} also endows the small world with a 
probabilistic interpretation.

In the following sections I review the measure-theoretic construction 
of frequentist and Bayesian inference.  The formality is a necessary evil
in order to identify the canonical measures of predictive performance that
do not rely on the additional structure of a particular measurement space.

\subsection{Frequentist Inference}

In frequentist inference \cite{VanDerVaart:1998, Rice:2006}
probabilities are defined strictly as frequencies of repeatable events, 
namely measurements.  Consequently probability theory applies to 
only the latent data generating process, $\pi$, and the data generating 
processes in the small world, $ \left\{ \pi_{x} \right\}$.  Considered as 
a family of probability distribution functions on the measurement space,
the small world, 
\begin{equation*}
\pi_{x} : X \times \mathcal{Y} \rightarrow \left[ 0, 1 \right],
\end{equation*}
is otherwise known as a \textit{likelihood function}.  

Given this rigid definition of probability, the only way we can construct a 
complete predictive distribution is by selecting a single element of the small 
world and utilizing the corresponding data generating process.  One of the 
most prolific approaches to selecting such an element is with the use of
\textit{estimators}, functions of the data that identify some aspect of the 
latent data generating process.  Formally, estimators are defined as maps 
from the measurement space to some auxiliary space, 
$\hat{e} : Y \rightarrow Z$.  

Given a set of estimators, $E$, we can quantify how well a given estimator 
identifies the latent data generating process with a \textit{loss function},
\begin{equation*}
L : Y \times E \rightarrow \mathbb{R}.
\end{equation*}
The corresponding \textit{risk} of an estimator is defined as
\begin{align*}
R :& \, X \times E \rightarrow \mathbb{R} \\
& \left( x, \hat{e} \right) \mapsto 
\int_{Y} \pi_{x} \! \left( \mathrm{d} y \right) L \! \left( y, \hat{e} \right),
\end{align*}
and \textit{minimax estimators} are defined by the optimality criterion,
\begin{equation*}
\hat{e}_{M} = 
\underset{\hat{e} \, \in \, E}{\mathrm{argmin}} \,
\underset{x \, \in \, X}{\mathrm{max}} \,
R \! \left(x, \hat{e} \right).
\end{equation*}
With a map $f: Z \rightarrow X$ we can then select a single element 
of the small world and define the subsequent predictive distribution 
for new data by
\begin{equation*}
\widetilde{\pi}_{Y | y} \equiv 
\pi_{f ( \hat{e}_{M} ( y ) )}.
\end{equation*}  

For example, consider the circumstance where the small world
contains the latent data generation process,
\begin{equation*}
\pi = x_{\pi} \in X,
\end{equation*}
and take any map $g : X \rightarrow Z$ with a well-defined inverse, 
$g^{-1} : g \! \left( X \right) \rightarrow X$.  If $Z$ is a metric space 
then a natural loss function is given by the distance function,
\begin{equation*}
L \! \left( y, \hat{e} \right) = 
D \! \left( g \! \left( x_{\pi} \right), \hat{e} \! \left( y \right) \right);
\end{equation*}
the resulting minimax estimator, $\hat{e}_{M}$, approximates the
true value of the function, $g \! \left( x_{\pi} \right)$, which then identifies
a unique element of the small world,
\begin{align*}
\hat{x} = g^{-1} \circ \hat{e}_{M}  :& \, Y \rightarrow X \\
& \; y \mapsto g^{-1} \! \left( \hat{e}_{M} \! \left( y \right) \right).
\end{align*}

\textit{Maximum likelihood estimators} avoid the need for a loss
function by using the likelihood itself.  Given any reference measure,
$\lambda$, with respect to which every element of the small
world is absolutely continuous, the maximum likelihood estimator
is defined as
\begin{align*}
\hat{x}_{\mathrm{MLE}} :&\, Y \rightarrow X \\
& y \mapsto 
\underset{x \, \in \, X}{\mathrm{argmax}}
\frac{ \mathrm{d} \pi_{x} }{  \mathrm{d} \lambda} \! \left( y \right).
\end{align*}
Provided that the maximum is unique, $\hat{x}_{\mathrm{MLE}}$
identifies a unique element of the small world and hence an
unambiguous predictive distribution.

The practical utility of a frequentist estimator is a subtle issue.  
When the small world does not contain the latent data generating 
process, for example, any predictive distribution derived from an
estimator will never be able to recover the latent data generating 
process exactly.  Moreover, even if the small world does contain the true 
data generating process there is no guarantee that an estimator
evaluated at a given measurement will identify it.  A given
estimator may be \textit{unidentified} and unable to select a single
element of the small world at all, or it may simply be inaccurate or
imprecise.  In any case we must be skeptical of how well 
$\widetilde{\pi}_{Y | y}$ approximates the latent data generating 
process.

\subsection{Bayesian Inference}

Bayesian inference \cite{BernardoEtAl:2009, GelmanEtAl:2014a} considers 
a more general interpretation of probability that encompasses not just 
frequencies but any self-consistent system of uncertainty.  Consequently we can 
assign probability distributions to not only the measurement space but also the 
small world itself given the choice of a $\sigma$-algebra, $\mathcal{X}$, on X.

From this perspective, the small world now defines a regular conditional 
probability distribution,
\begin{equation*}
\pi_{Y | x} : \mathcal{Y} \times X \rightarrow [0, 1],
\end{equation*}
with respect to the canonical projection operator,
\begin{equation*}
\varpi_{X} : Y \times X \rightarrow X,
\end{equation*}
on the product space of measurements and the small world, $Y \times X$.
The difference between this regular conditional probability distribution
and the frequentist likelihood function is largely one of interpretation and, 
following convention, I will refer to both objects as likelihoods.

Inference proceeds with the introduction of a \textit{prior distribution}
over the small world,
\begin{equation*}
\pi_{X} : \mathcal{X} \rightarrow [0, 1],
\end{equation*}
that encodes all information about the latent data generating process within
the context of the small world before the current measurement is made.  Together 
with the likelihood, the prior distribution defines a joint distribution on the product 
space of measurements and the small world, $\pi_{Y \times X}$, and information 
about the small world given a measurement is encoded in the regular conditional 
probability distribution,
\begin{equation*}
\pi_{X | y} : \mathcal{X} \times Y \rightarrow [0, 1],
\end{equation*}
defined with respect to  the second canonical projection operator,
\begin{equation*}
\varpi_{Y} : Y \times X
\rightarrow Y.
\end{equation*}

Conditioning $\pi_{X | y}$ on a given measurement, $y$, gives the
\textit{posterior distribution} over the small world,
\begin{equation*}
\pi_{X | y} = 
\left[ \frac{ \mathrm{d} \pi_{Y | x} }
{ \mathrm{d} \left( \varpi_{Y} \right)_{*} \pi_{Y \times X} } \! \left( y \right) \right]
\, \pi_{X}.
\end{equation*}
When $Y \sim X \sim \mathbb{R}^{n}$, the posterior density with respect to 
a reference Lebesgue measure is given by the celebrated Bayes' Rule,
\begin{equation*}
\pi_{X | y} \! \left( x | y \right)
= 
\frac{ \pi_{Y | x} \! \left( y | x \right) \pi_{X} \! \left( x \right) }
{ \int_{\mathbb{R}^{n}} \mathrm{d} x \, \pi_{Y | x} \! \left( y | x \right) 
\pi_{X} \! \left( x \right) }.
\end{equation*}

Given the more general application of probability in Bayesian inference
we can construct a predictive distribution not only by selecting a single
element of the small world but also by averaging the elements of the
small world with respect to a given probability distribution.  The 
\textit{prior predictive distribution}, for example, is given by weighting 
each element of the small world according to the prior distribution,
\begin{equation*}
\pi_{Y}^{\mathrm{prior}} \! \left( \mathrm{d} \tilde{y} \right) = 
\int_{X} \pi_{X} \! 
\left( \mathrm{d} x \right) \pi_{Y | x} \! \left( \mathrm{d} \tilde{y}, x \right).
\end{equation*}
Similarly, the \textit{posterior predictive distribution} is given by
weighting each element of the small world according to the posterior 
distribution,
\begin{equation*}
\pi_{Y | y}^{\mathrm{post}} \! \left( \mathrm{d} \tilde{y} | y \right) = 
\int_{X} \pi_{X | y} \! \left( \mathrm{d} x | y \right) 
\pi_{Y | X} \! \left( \mathrm{d} \tilde{y} | x \right).
\end{equation*}
Because it learns from the measured data, the posterior
predictive distribution should be a better approximation of the
latent data generating process provided that the modeling
assumptions, such as the choice of the small world and the
prior distribution, are compatible with the true data generating
process.

As in frequentist inference, the performance of either predictive 
distribution depends critically on the assumptions in their
construction.  Some means of comparing the chosen predictive 
distribution to the latent data generating process is vital for 
validating the modeling assumptions and ensuring inferences 
that perform well in practice.

\section{Validating Inference}

Although the frequentist and Bayesian approaches have different
means of inferring predictive distributions, they can both succumb
to the same pathologies that jeopardize predictive performance.

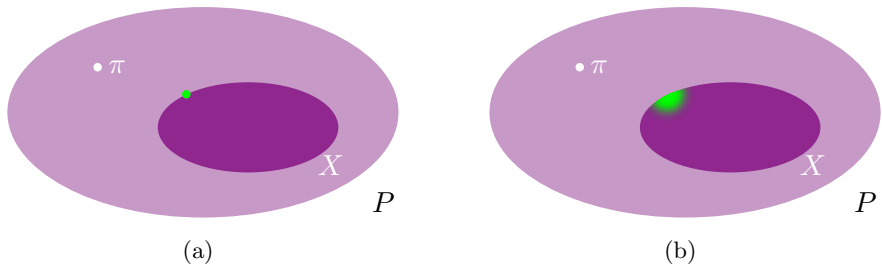
\begin{figure*}
\centering
\subfigure[]{
\begin{tikzpicture}[scale=0.20, thick]

  \draw[color=white] (-15, 0) -- (15, 0);

  \fill[light] (0, 0) ellipse (13 and 7);
  \node at (12, -6) {$P$};
  
  \fill[dark] (3, -1) ellipse (6 and 3);
  \node[color=white] at (8.5, -3.5) {$X$};
  
  \fill[color=white] (-7, 3) circle (8pt)
  node[right, color=white] {$\pi$};
  
  \fill[color=green] (-1.1, 1.2) circle (8pt);
  
\end{tikzpicture}
}
\subfigure[]{
\begin{tikzpicture}[scale=0.20, thick]

  \draw[color=white] (-15, 0) -- (15, 0);

  \fill[light] (0, 0) ellipse (13 and 7);
  \node at (12, -6) {$P$};
  
  \fill[dark] (3, -1) ellipse (6 and 3);
  \node[color=white] at (8.5, -3.5) {$X$};
  
  \fill[color=white] (-7, 3) circle (8pt)
  node[right, color=white] {$\pi$};
  
  \begin{scope}
    \clip (3, -1) ellipse (6 and 3);
    \foreach \i in {0, 0.05,..., 1} {
      \fill[opacity={exp(-5 * \i*\i)}, green] (-1.2, 1.2) circle ({2 * \i});      
    }
  \end{scope}
  
\end{tikzpicture}
}
\caption{Because (a) frequentist estimators and (b) Bayesian
prior and posterior distributions are limited to the small world,
neither approach will be able to  construct a predictive distribution
capable of exactly modeling the latent data generating process, $\pi$,
when it is not an element of the small world.
}
\label{fig:misfit}
\end{figure*}

\begin{figure*}
\centering
\subfigure[]{
\begin{tikzpicture}[scale=0.20, thick]
  
  \draw[color=white] (-15, 0) -- (15, 0);
  
  \fill[light] (0, 0) ellipse (13 and 7);
  \node at (12, -6) {$P$};
  
  \fill[dark] (-4, 2) ellipse (6 and 3);
  \node[color=white] at (1.5, -0.5) {$X$};
  
  \fill[color=white] (-7, 3) circle (8pt)
  node[right, color=white] {$\pi$};
  
  \fill[color=green] (-3.5, 1.5) circle (8pt);
  
\end{tikzpicture}
}
\subfigure[]{
\begin{tikzpicture}[scale=0.20, thick]
  
  \draw[color=white] (-15, 0) -- (15, 0);
  
  \fill[light] (0, 0) ellipse (13 and 7);
  \node at (12, -6) {$P$};
  
  \fill[dark] (-4, 2) ellipse (6 and 3);
  \node[color=white] at (1.5, -0.5) {$X$};
  
  \fill[color=white] (-7, 3) circle (8pt)
  node[right, color=white] {$\pi$};

  \foreach \i in {0, 0.05,..., 1} {
    \fill[opacity={exp(-5 * \i*\i)}, green] (-3.5, 1.5) circle ({2 * \i});      
  }
  
\end{tikzpicture}
}
\caption{Even when the small world does contain the latent data
generating process, $\pi$, inferences are not guaranteed to capture 
it.  Here (a) a frequentist estimator evaluated at a given measurement
strays from $\pi$ while (b) a Bayesian prior or posterior
distribution concentrates away from $\pi$.  In either case the
predictive distributions will be biased away from the latent data
generating process.
}
\label{fig:overfit}
\end{figure*}
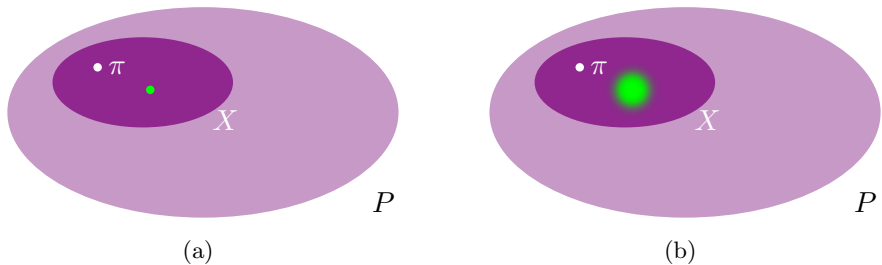

The most obvious pathology is model \textit{misfit} where the true
data generating process is not contained within the small world,
$\pi \notin X$, and any inferential method will be able to 
approximate the exact predictive distribution only so well 
(Figure \ref{fig:misfit}).  Even if the small world contains the latent data 
generating process, however, inferences still may not be able to find 
it because they \textit{overfit} to irrelevant structure in the measurement, 
such as purely stochastic noise (Figure \ref{fig:overfit}).  In practice
these two pathologies are somewhat antagonistic -- making a model 
more complex in order to reduce misfit often renders it more vulnerable 
to noise and hence subject to overfitting.

\begin{figure*}
\centering
\subfigure[]{\includegraphics[width=1.9in]{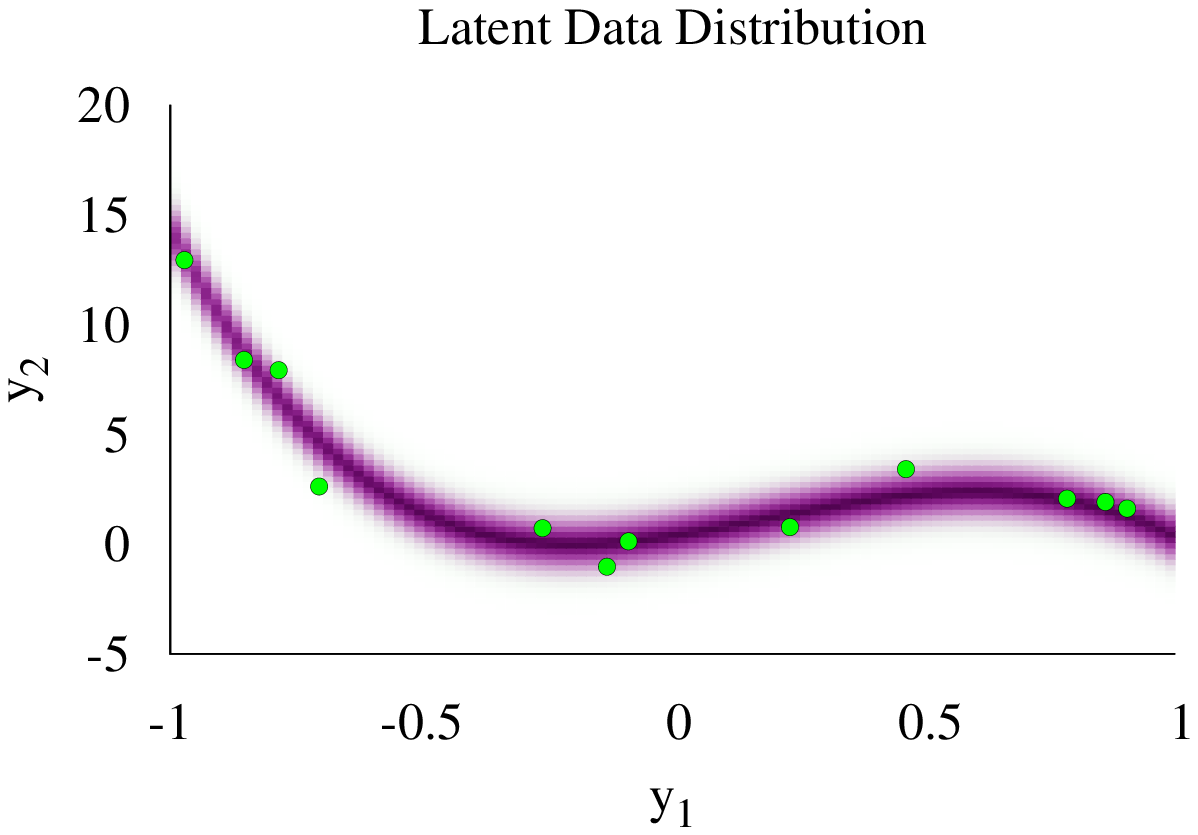}}
\subfigure[]{\includegraphics[width=1.9in]{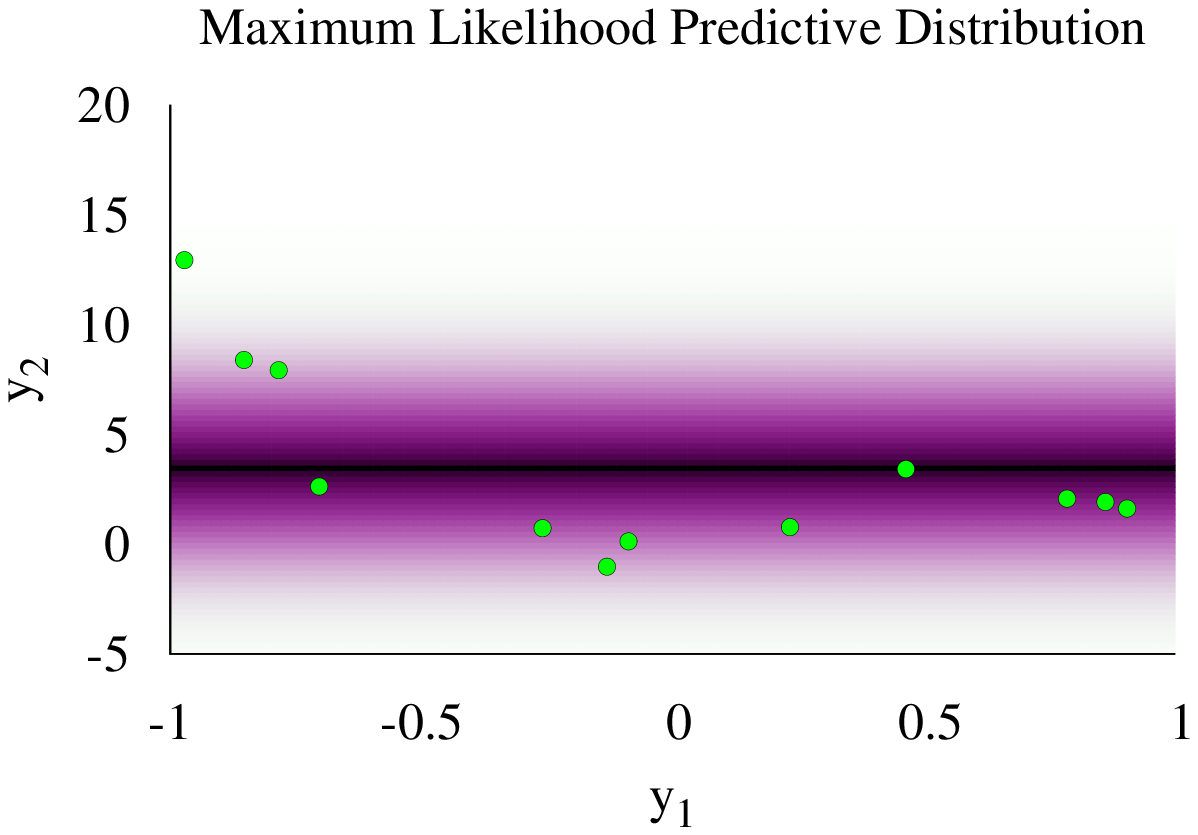}}
\subfigure[]{\includegraphics[width=1.9in]{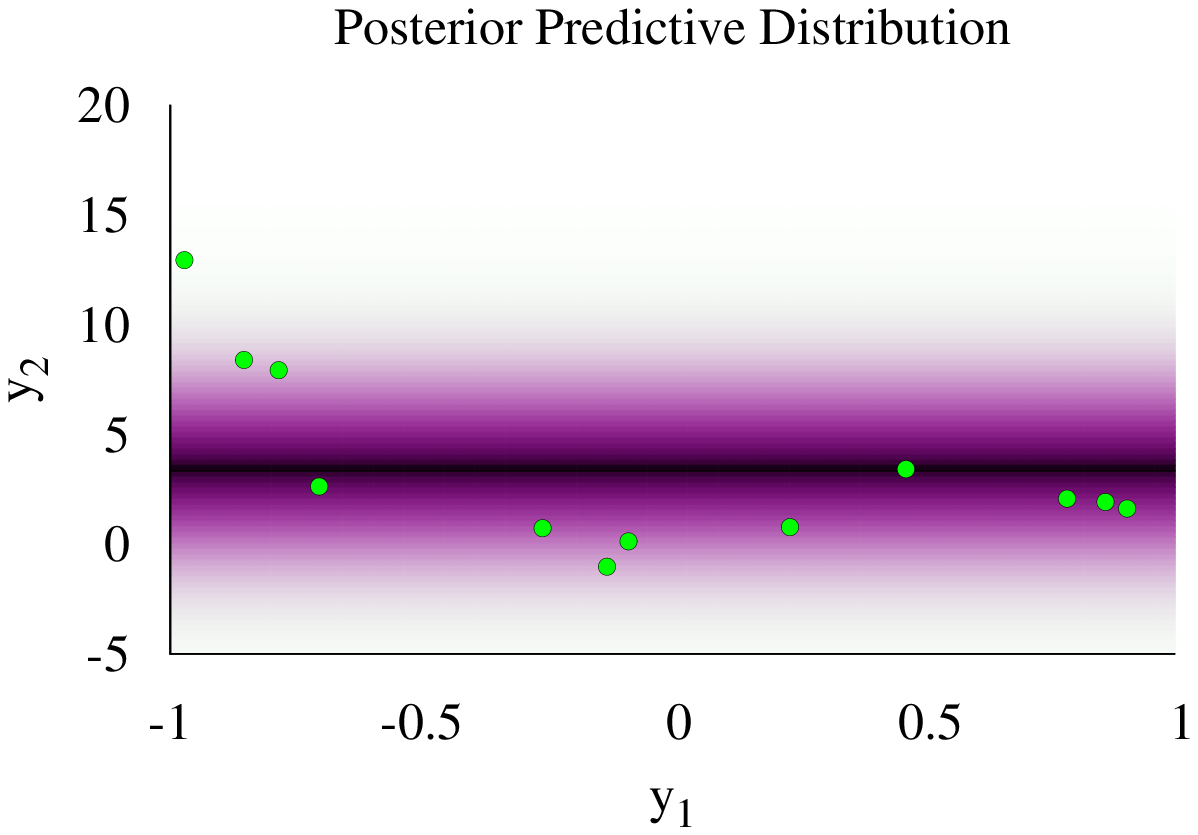}}
\caption{Misfit occurs when an inferential model cannot capture the 
complexity of the latent data generating process, for example when 
models assuming the simple data generating process \eqref{eqn:simple_data} 
are fit to (a) data generated according to the more complex process 
\eqref{eqn:complex_data}.  Because the assumptions are too restrictive, 
the resulting (b) maximum likelihood predictive distribution and (c) posterior 
predictive distribution are poor approximations to the latent data generating
process.  Compare to Figure \ref{fig:misfit}.
}
\label{fig:mistfit_example}
\end{figure*}

\begin{figure*}
\centering
\subfigure[]{\includegraphics[width=1.9in]{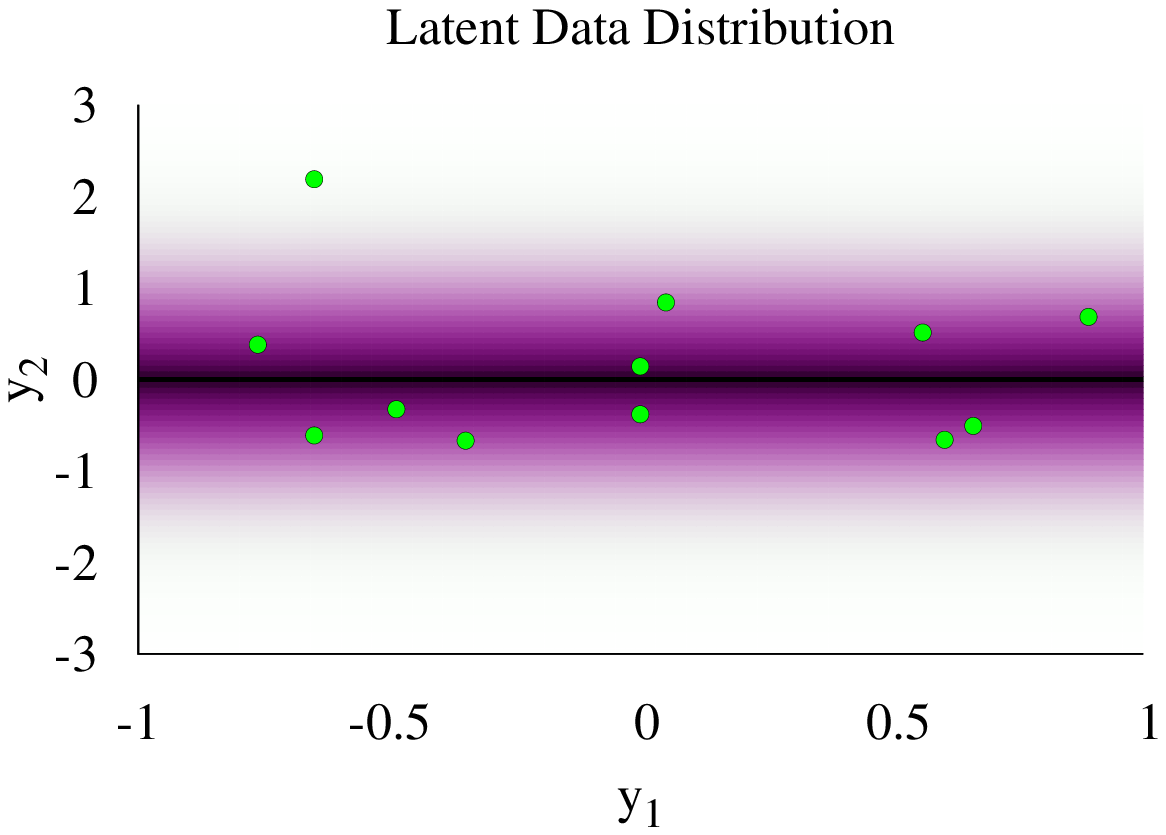}}
\subfigure[]{\includegraphics[width=1.9in]{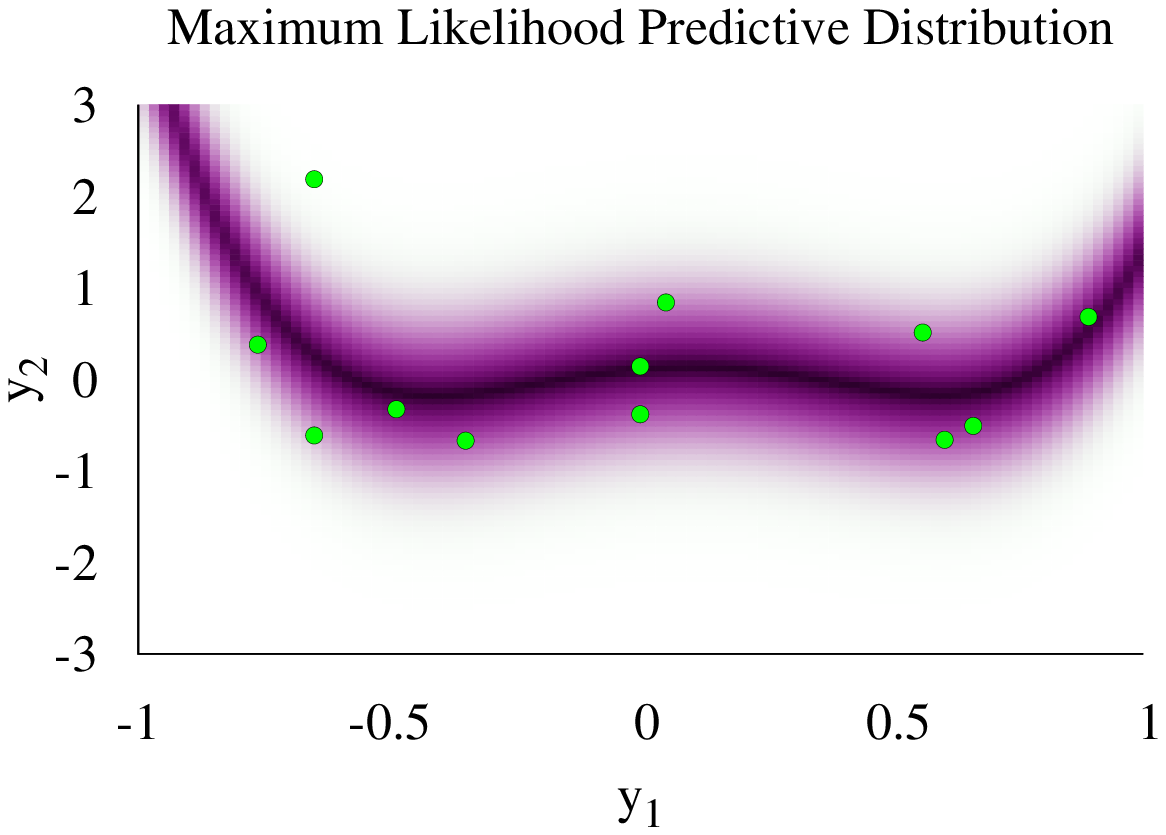}}
\subfigure[]{\includegraphics[width=1.9in]{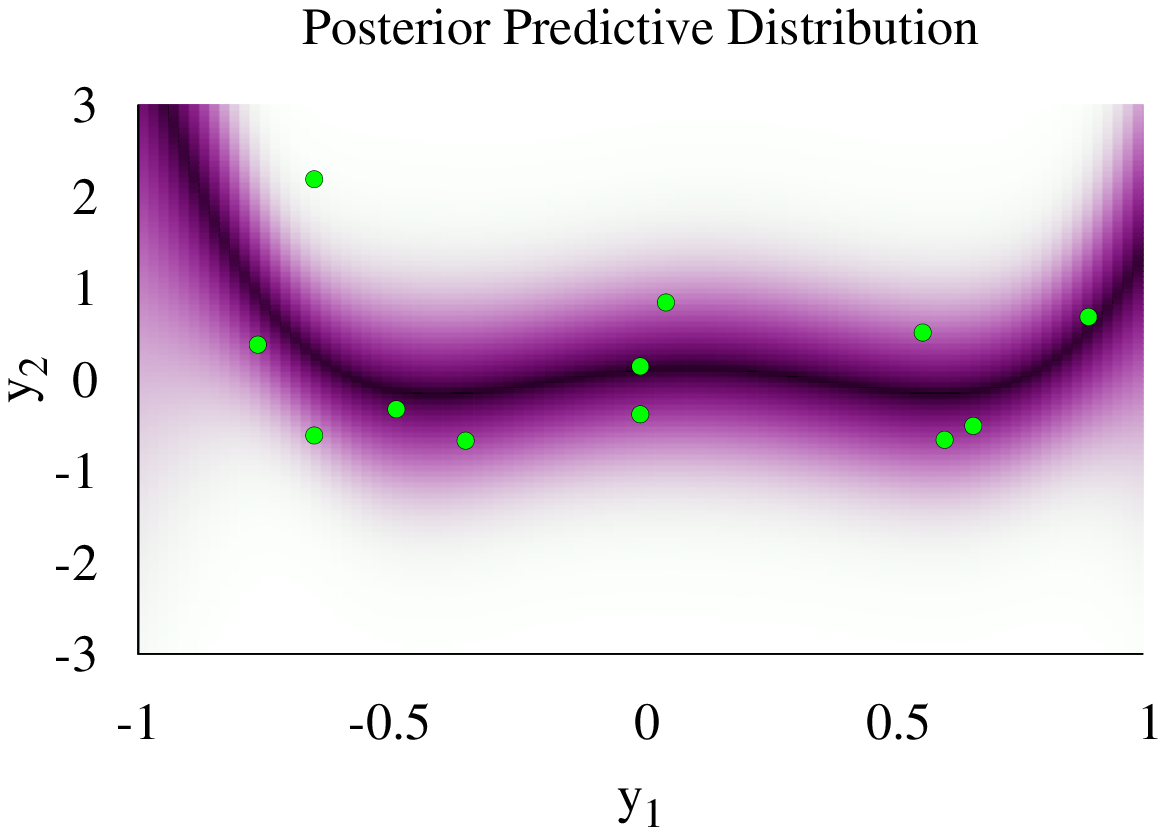}}
\caption{Overfitting occurs when an inferential model has too much
flexibility relative to the latent data generating process, for example
when models assuming a complex data generating process 
\eqref{eqn:complex_data} are fit to (a) data generating according to
the simpler data generating process \eqref{eqn:simple_data}.  Both
the (b) maximum likelihood predictive distribution and (c) posterior
predictive distribution recklessly fit to the Gaussian noise in the data, 
biasing predictions away from the latent data generating process.  
Compare to Figure \ref{fig:overfit}.
}
\label{fig:overfit_example}
\end{figure*}

Consider, for example, the measurement space 
$Y = \left( \mathbb{R} \times \mathbb{R} \right)^{N = 12}$ with two data 
generating processes: a Gaussian distribution centered on a quartic polynomial 
\begin{equation} \label{eqn:complex_data}
y_{1,n} \sim U \! \left(-1, 1 \right), \,
y_{2,n} | y_{1,n} \sim \mathcal{N} \! \left( \sum_{k = 0}^{4} c_{k} y^{k}_{1, n}, \sigma^{2} \right), \,
n = 1, \ldots, 12,
\end{equation}
and a Gaussian distribution centered on a constant,
\begin{equation} \label{eqn:simple_data}
y_{1,n} \sim U \! \left(-1, 1 \right), \,
y_{2,n} \sim \mathcal{N} \! \left( c_{0}, \sigma^{2} \right), \,
n = 1, \ldots, 12.
\end{equation}
Misfit occurs when the measurement is generated from the more 
complex process \eqref{eqn:complex_data} (Figure 
\ref{fig:mistfit_example}a) but the inferential models assume the simpler 
process \eqref{eqn:simple_data}; in this case the resulting predictive 
distributions (Figure \ref{fig:mistfit_example}b, c) will never be able 
to capture the latent data generating process.  On the other hand, 
overfit occurs when the measurement is generating from the simpler 
process (Figure \ref{fig:overfit_example}a) but the inferential models 
assume the more complex process.  The resulting predictive distributions 
(Figure \ref{fig:overfit_example}b, c) will overfit to the Gaussian noise, 
inducing a bias away from the latent data generating process.  In both 
cases the Bayesian analysis uses the conjugate prior
\begin{equation*}
\pi_{X} \! \left( \mathbf{c}, \sigma \right) 
= \mathrm{MultiNormalGamma} \! \left( 
\mbox{\boldmath{$\mu$}}_{0}, \mathbf{\Lambda}_{0},
\alpha_{0}, \beta_{0} \right)
\end{equation*}
with
\begin{align*}
\mbox{\boldmath{$\mu$}}_{0} &= \mathbf{0}
\\
\mathbf{\Lambda}_{0} &= 0.001 \cdot \mathbb{I}
\\
\alpha_{0} &= 0.5
\\
\beta_{0} &= 0.5.
\end{align*}

Reliable inferences consequently require some predictive validation
to ensure that, even if the model misfits or overfits, the resulting predictive 
distribution approximates the latent data generating process sufficiently 
well.  One immediate strategy is to test the model within a null hypothesis 
significance testing framework, rejecting if the measured data is sufficiently 
unlikely with respect to the inferred predictive distribution.  By construction, 
however, we make no attempt to model anything outside of the small world, 
let alone its entire complement, which prevents us from constructing a valid 
alternative hypothesis needed to calibrate such tests.  In order to quantify 
predictive performance without looking outside of the small world we need 
to compare the predictive distribution to the latent data generating process 
directly.

\cite{VehtariEtAl:2012} considered many possible strategies for
comparing predictive distributions to the latent data generating process, 
but almost all of them require endowing the small world with additional 
structure, such as a metric or a distinguished test function, that limits
the ultimate scope of the validation.  Only one of the approaches 
considered arises canonically from the general construction of 
inference -- the Kullback-Leibler divergence \cite{KullbackEtAl:1951}.  
In this section I discuss how the Kullback-Leibler divergence defines a
measure of relative predictive performance, although one that cannot 
be computed in practice.  I then consider a manipulation of the 
Kullback-Leibler divergence that defines scores of relative predictive 
performance that are amenable to approximations, and finally I show 
how various approximation strategies give rise to many of the model
comparison techniques already popular in practice.

\subsection{Relative Predictive Performance Measures}

In order to construct a measure of predictive performance we have to
compare some inferred predictive distribution, $\widetilde{\pi}_{Y | y}$,
to the latent data generating process, $\pi$.  Without endowing the
measurement space with any particular structure, the only canonical
way of comparing two distributions, $\mu$ and $\nu$, on $Y$ is with 
an $f$-divergence \cite{CsiszarEtAl:2004},
\begin{equation*}
D_{f} \! \left( \mu \mid\mid \nu \right) 
= \int_{Y} \nu \! \left( \mathrm{d} y \right)
f \! \left( \frac{ \mathrm{d} \mu }{ \mathrm{d} \nu } \! \left( y \right) \right),
\end{equation*}
where $f : \mathbb{R} \rightarrow \mathbb{R}$ is any convex function
satisfying $f \! \left( 1 \right) = 0$.  Moreover, the only $f$-divergence 
that respects any product structure of the measurement space and 
allows us to marginalize out irrelevant structure as necessary is the 
Kullback-Leibler divergence,
\begin{equation*}
\mathrm{KL} \! \left( \mu \mid\mid \nu \right) 
= - \int_{Y} \mu \! \left( \mathrm{d} x \right)
\log \frac{ \mathrm{d} \nu }{ \mathrm{d} \mu } \! \left( x \right).
\end{equation*}
The Kullback-Leibler divergence vanishes only when the two measures 
are equal and monotonically increases as the two measures deviate, 
approaching infinity when $\nu$ is not absolutely-continuous with 
respect to $\mu$.

Because the Kullback-Leibler divergence is not symmetric there are
two possible ways that we might use it to compare the latent data
generating process and an inferred predictive distribution.  Using the 
inferred predictive distribution as the base measure, 
$\mathrm{KL} \! \left( \widetilde{\pi}_{Y | y} \mid\mid \pi \right) $,
considers the predictive performance only where $\widetilde{\pi}_{Y | y}$
concentrates and consequently does not penalize predictive distributions 
that completely ignore neighborhoods supported by $\pi$.  In
an extreme limit, this divergence does not even penalize predictive
distributions that are not absolutely continuous with respect to the
latent data generating process.  In order to truly assess the inferred
predictive distribution we need to instead base the divergence on the 
latent data generating process itself, 
\begin{equation*}
\mathrm{KL} \! \left(  \pi \mid\mid \widetilde{\pi}_{Y | y} \right)
= - \int_{Y} \pi \! \left( \mathrm{d} \tilde{y} \right)
\log \frac{ \mathrm{d} \widetilde{\pi}_{Y | y} }
{ \mathrm{d} \pi } \! \left( \tilde{y} \right).
\end{equation*}
Here we will use this form of the Kullback-Leibler divergence to
quantify the validity of our modeling assumptions, but it can also be used
to construct a more elaborate sensitivity analysis of those assumptions
\cite{WatsonEtAl:2014}.

As with null hypothesis significance testing, the Kullback-Leibler divergence
cannot be calibrated, in other words there is no canonical threshold below 
which we can declare our model assumptions valid.  Unlike hypothesis 
testing, however, the difference between two divergences is meaningful,
allowing us to quantify the \textit{relative} performance of 
$\widetilde{\pi}_{Y | y}$ compared to some other predictive distribution.  
Although $\mathrm{KL} \! \left( \pi \mid\mid \widetilde{\pi}_{Y | y} \right)$ cannot
be computed without assuming a priori knowledge of the true data generation 
process, we can manipulate the divergence into a more advantageous form 
without compromising its quantification of relative performance.

Let $\lambda$ be any reference measure with respect to which both the true 
data generating process and the inferred predictive distribution are absolutely 
continuous.  We can then define a \textit{relative predictive performance score} 
as
\begin{align} \label{rpps}
\delta \! \left( \pi \mid\mid \widetilde{\pi}_{Y | y} \right)
&= \mathrm{KL} \! \left( \pi \mid\mid \widetilde{\pi}_{Y | y} \right) 
- \mathrm{KL} \! \left( \pi \mid\mid \lambda \right)
\\
&= 
- \int_{Y} \pi \! \left( \mathrm{d} \tilde{y} \right)
\log \frac{ \mathrm{d} \widetilde{\pi}_{Y | y} }
{ \mathrm{d} \lambda } \! \left( \tilde{y} \right) \nonumber
\\
&= 
- \mathbb{E}_{\pi} \! \left[
\log \frac{ \mathrm{d} \widetilde{\pi}_{Y | y} }
{ \mathrm{d} \lambda } \right]. \nonumber
\end{align}
The difference between any relative predictive performance scores is the
same as the difference of the equivalent Kullback-Leibler divergences
and so they quantify the same relative performance, but because the 
densities $\mathrm{d} \widetilde{\pi}_{Y | y} / \mathrm{d} \lambda$
are independent of the latent data generating process these relative scores
can be approximated using only sampled measurements from $\pi$.  
Relative predictive performance scores also have the welcome interpretation 
as expected logarithmic score functions \cite{GneitingEtAl:2007, BernardoEtAl:2009}.

The ultimate utility of these relative predictive performance scores then
depends on the accuracy and precision of the chosen approximation strategy.

\subsection{Approximating Relative Predictive Performance Scores}

Although relative predictive performance scores cannot be calculated 
exactly, their construction makes them amenable to a variety of
approximations.  For example, given an ensemble of $N + 1$ 
measurements we could construct a Monte Carlo estimator,
\begin{equation*}
\hat{\delta} \! \left( \pi \mid\mid \widetilde{\pi}_{Y | y} \right)
\approx
- \frac{1}{N} \sum_{n = 1}^{N}
\log \frac{ \mathrm{d} \widetilde{\pi}_{Y | y_{N + 1}} }{ \mathrm{d} \lambda } 
\! \left( y_{n} \right),
\end{equation*}
with vanishing bias and quantifiable variance.  Unfortunately, in practice 
we rarely have an ensemble of measurements and instead have to 
consider approximations that utilize only a single measurement. 

\subsubsection{Delta Estimators}

An immediate approximation of relative predictive performance 
scores derives from making a delta approximation of the latent
data generating process, $\pi \approx \delta_{y}$, which gives
\begin{equation*}
\hat{\delta}_{D} \! \left( \pi \mid\mid \widetilde{\pi}_{Y | y} \right)
\equiv
- \log \frac{ \mathrm{d} \widetilde{\pi}_{Y | y} }{ \mathrm{d} \lambda } 
\! \left( y \right).
\end{equation*}

Using the same measurement to learn the model and then validate it 
introduces a bias that makes delta estimators susceptible to overfitting.
Moreover, the underlying delta approximation typically induces a large
variance in the estimator, making fine comparisons between models 
difficult if not impossible.

\subsubsection{Hold-out Estimators}

More sophisticated estimates of relative predictive performance 
scores can be constructed by using the given measurement to
simulate an ensemble of measurements.  

Assuming that the measurement space has a product
structure, $Y = \prod_{n = 1}^{N} Y_{n}$, any measurement
can be partitioned in two subsets of size $N_{1}$ and $N_{2}$.  
Hold-out estimators use one of these partitions, often denoted the 
\textit{training data}, to infer a predictive distribution and the remaining 
partition, denoted the \textit{validation data}, to construct a delta 
estimator,
\begin{align*}
\hat{\delta}_{H} \! \left( \pi \mid\mid \widetilde{\pi}_{Y | y} \right)
&\equiv
- \log \left( 
\frac{ \mathrm{d} \widetilde{\pi}_{Y | y_{1}} }{ \mathrm{d} \lambda } \! \left( y_{2} \right) 
\right)^{N / N_{2}}
\\
&=
- \frac{N}{N_{2}}
\log \frac{ \mathrm{d} \widetilde{\pi}_{Y | y_{1}} }{ \mathrm{d} \lambda } 
\! \left( y_{2} \right).
\end{align*}

The simulated partitions in the validation data not only promise a more
precise estimate but also admit the estimation of the estimator variance
using the Monte Carlo standard error.  This, however, comes with the 
assumption the naive scaling of the predictive density inferred from 
the training data is a reasonable approximation to the predictive density 
inferred from the full measurement.  When data are sparse relative to the 
model complexity this assumption can severely bias the estimator; for 
example, predictive distributions inferred from small partitions are more 
susceptible to overfitting, artificially penalizing the predictive performance 
of model.

Moreover, the product structure of the measurement space necessary 
to construct hold-out estimators precludes many structured measurements,
such as those arising from some hierarchical models, networks, and time
series.

\subsubsection{Jackknife Estimators}

In order to compensate for the some of the potential bias in hold-out 
estimators we can appeal to a jackknife estimator \cite{Miller:1974}
which averages over the possible assignments of training and validation 
data.  Partitioning the measurement into $K$ subsets of size $M = N / K$,
the jackknife estimator is given by
\begin{align*}
\hat{\delta}_{J} \! \left( \pi \mid\mid \widetilde{\pi}_{Y | y} \right)
&\equiv
\frac{1}{K} \sum_{k = 1}^{K}
- \log \left( 
\frac{ \mathrm{d} \widetilde{\pi}_{Y | y \setminus y_{k} } }{ \mathrm{d} \lambda } 
\! \left( y_{k} \right) \right)^{\frac{N}{M}}
\\ 
&=
\frac{1}{K} \sum_{k = 1}^{K}
- \log \left( 
\frac{ \mathrm{d} \widetilde{\pi}_{Y | y \setminus y_{k} } }{ \mathrm{d} \lambda } 
\! \left( y_{k} \right) \right)^{K}
\\ 
&=
- \sum_{k = 1}^{K}
\log \frac{ \mathrm{d} \widetilde{\pi}_{Y | y \setminus y_{k} } }{ \mathrm{d} \lambda } 
\! \left( y_{k} \right),
\end{align*}
where $y \setminus y_{k}$ are often denoted the $k$th \textit{training data} and 
$y_{k}$ the $k$th \textit{testing data}.  This approach can also be readily generalized 
to a bootstrap estimator \cite{Efron:1979} which samples training and validation 
data with replacement.

The averaging over partitions typically reduces the bias of the relative predictive 
performance score estimation but the variance can still be quite large.
Moreover, the $K$ fits required to construct the jackknife estimator can
be prohibitively expensive in practice.

\subsection{Constructing Relative Predictive Performance Measures}

When we apply these approximation strategies to the predictive distributions 
arising in frequentist and Bayesian inference we immediately recover many 
of the comparative methods that have arisen and proved empirically effective 
in statistical practice.  In this section I detail many of these methods to 
emphasize the unifying nature of this foundational perspective.

\subsubsection{Comparing Likelihoods} \label{sec:comp_like}

In frequentist methods the inferred predictive distribution is given
by a single element in the small world or, equivalently, evaluating
the likelihood at a single point.

Explicit use of delta estimators of likelihood-based relative predictive 
performance scores provide a formal justification of the visual residual 
analysis \cite{AitchisonEtAl:1975, Tukey:1977} ubiquitous in not only 
statistics but also the physical sciences.  Moreover, when augmented
with an appropriate complexity penalty the the reuse estimator reduces 
to the Akaike Information Criterion \cite{Akaike:1971}.

Hold-out and jackknife estimators of likelihood-based relative 
predictive performance scores immediately yields predictive
log loss hold-out validation and cross validation, respectively,
which have become almost fundamental principles in the practice 
of modern machine learning \cite{Bishop:2006, HastieEtAl:2009}.

\begin{figure*}
\centering
\subfigure[]{\includegraphics[width=2.9in]{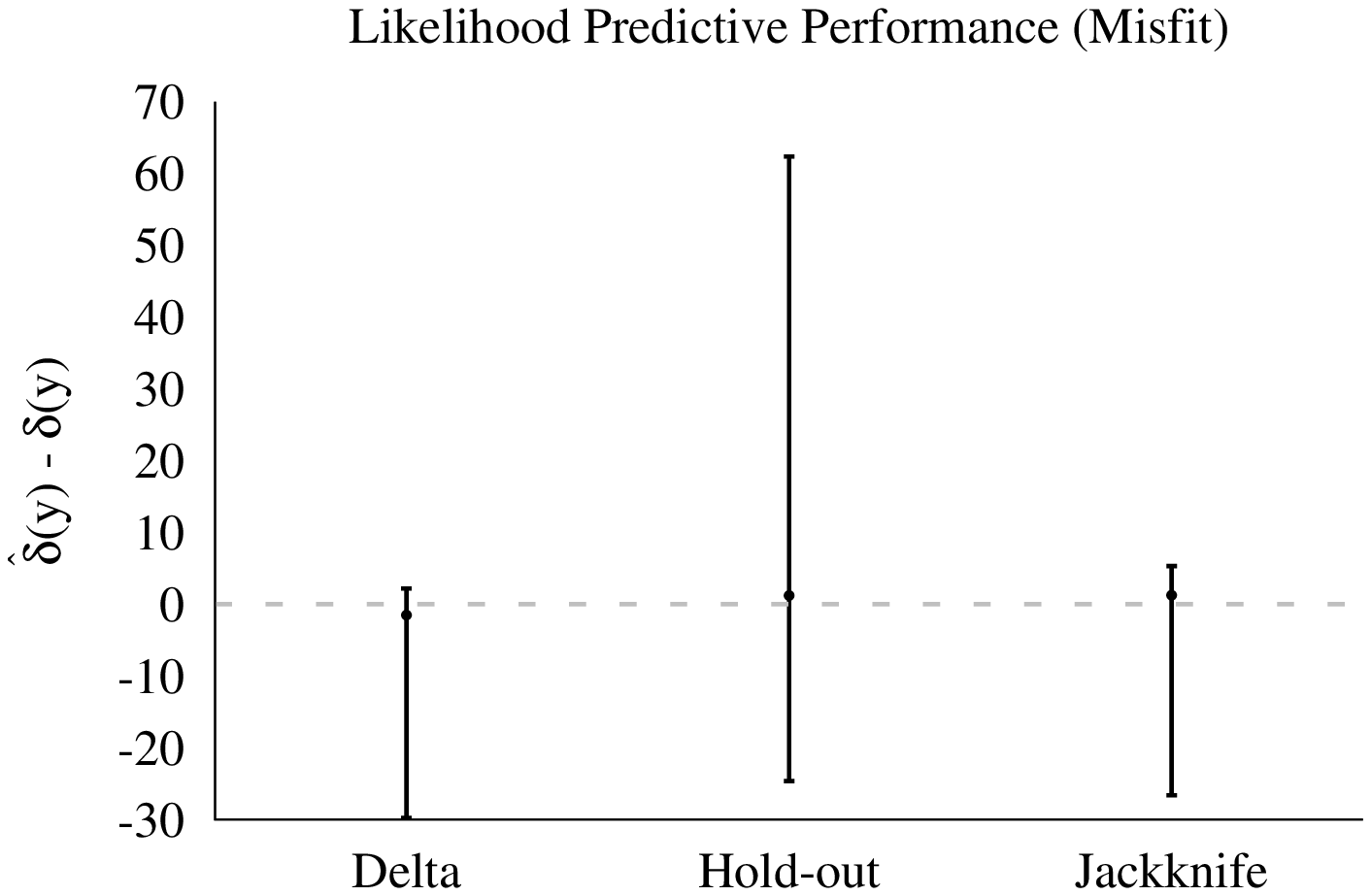}}
\subfigure[]{\includegraphics[width=2.9in]{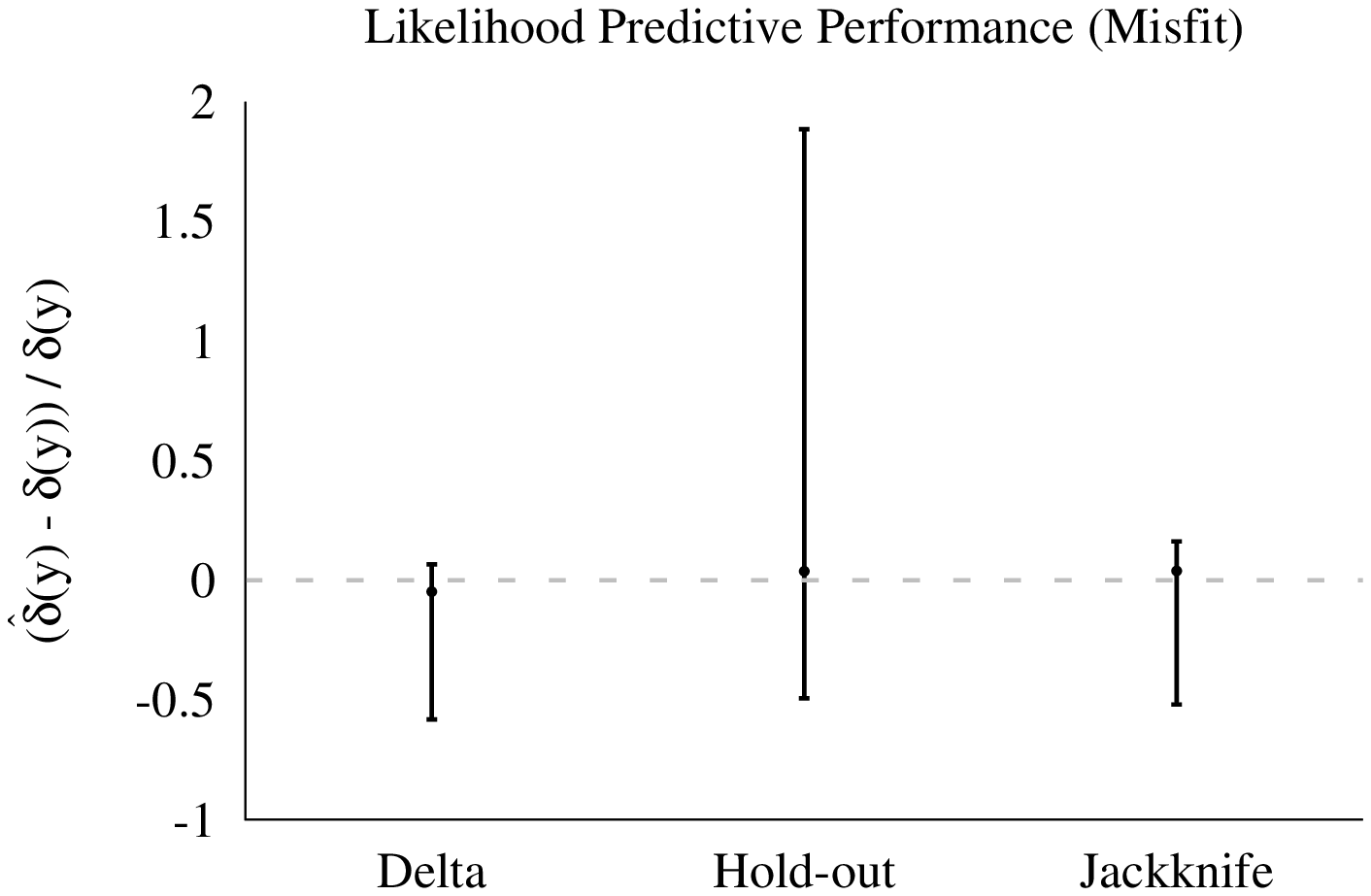}}
\caption{Even in the simple misfit model, approximations of 
likelihood-based relative predictive performance scores can leave 
much to be desired, as demonstrated by the $20\%$, $50\%$, and 
$80\%$ quantiles of the estimator error over an ensemble of 
measurements from the latent data generating process.
}
\label{fig:freq_pred_perf_misfit}
\end{figure*}

\begin{figure*}
\centering
\subfigure[]{\includegraphics[width=2.9in]{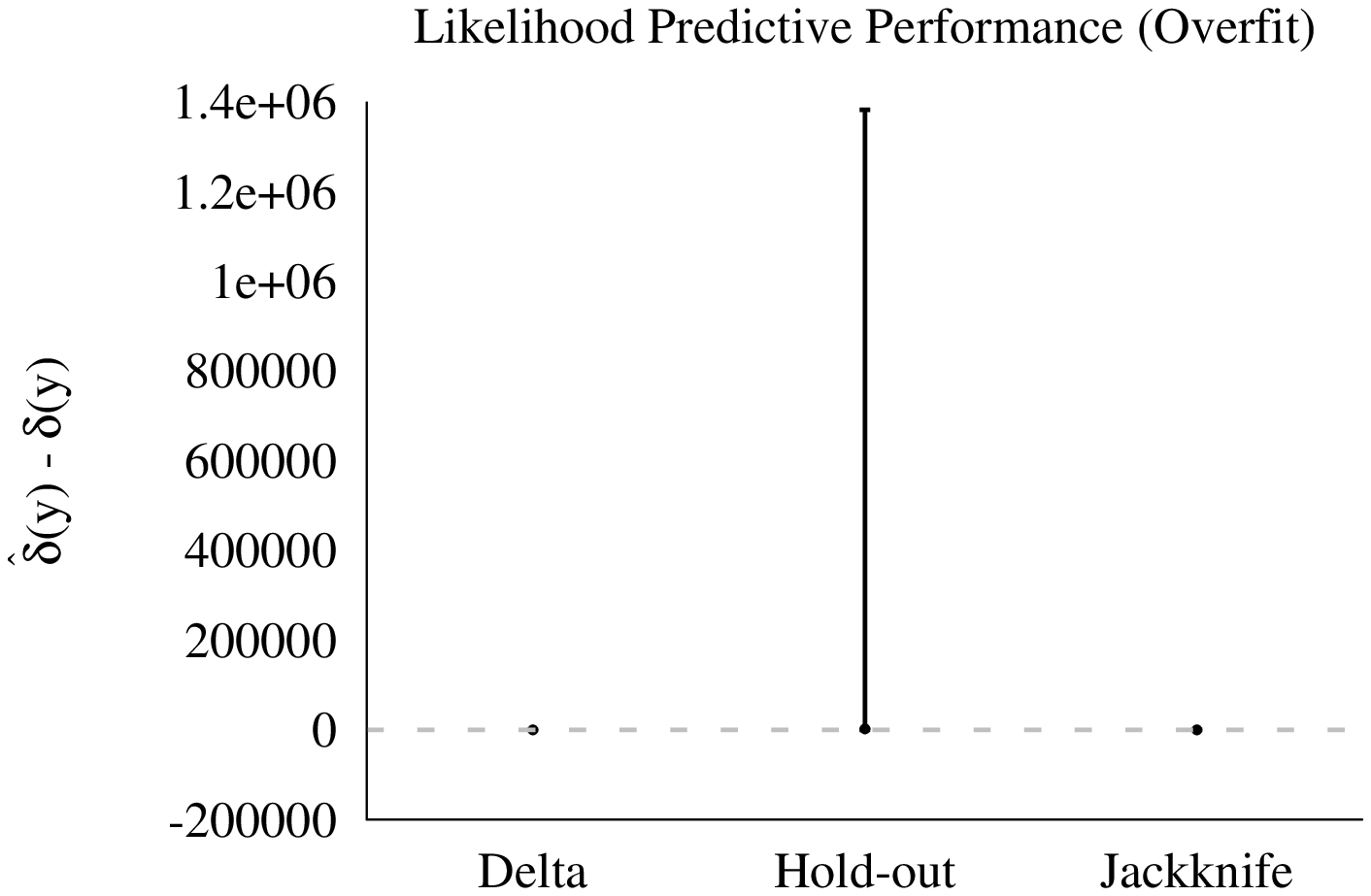}}
\subfigure[]{\includegraphics[width=2.9in]{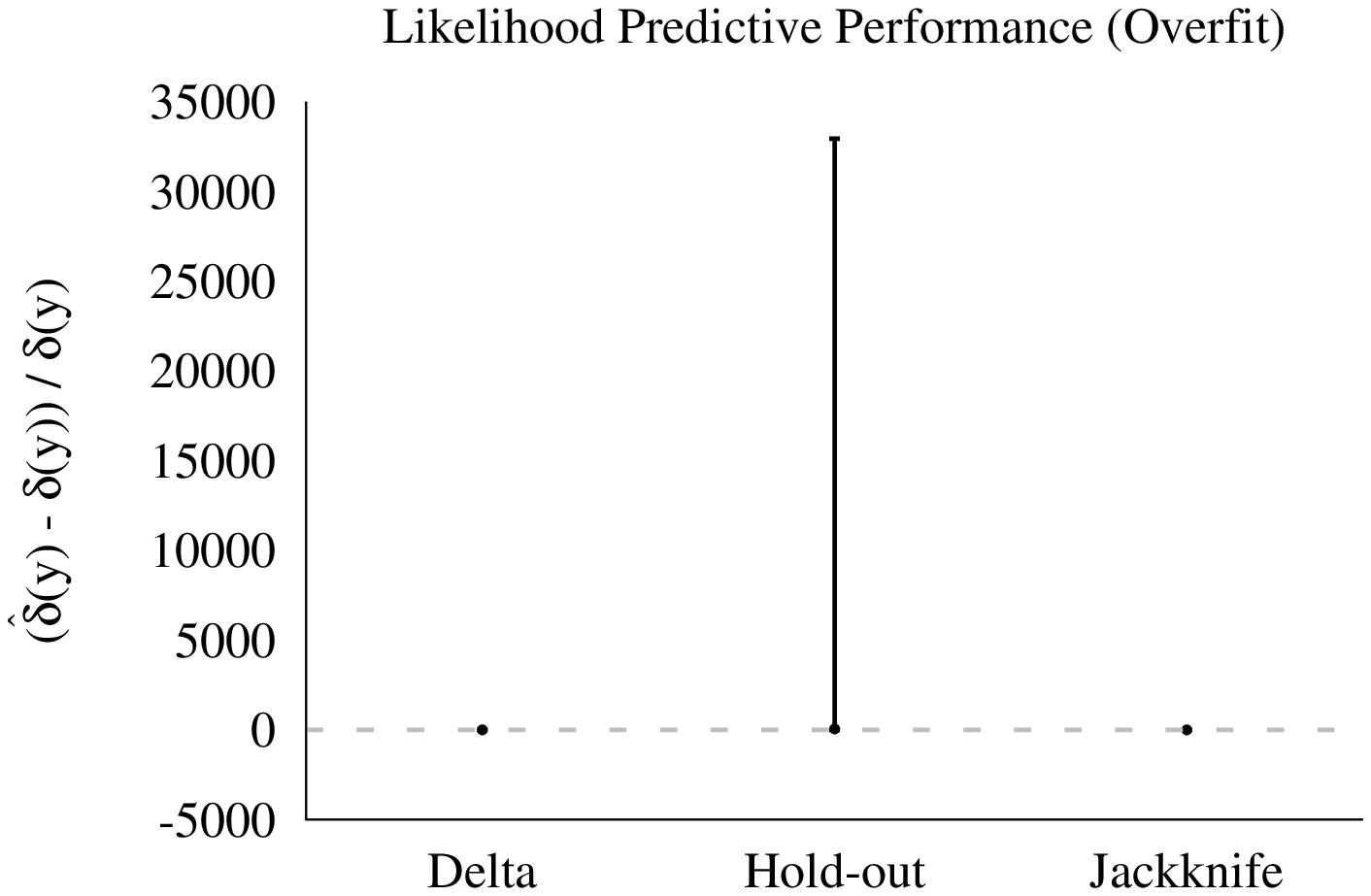}}
\caption{The simple overfit model also exposes the weakness of 
approximations of likelihood-based relative predictive performance 
scores can leave much to be desired, as demonstrated by the 
$20\%$, $50\%$, and $80\%$ quantiles of the estimator error over 
an ensemble of measurements from the latent data generating process.
Hold-out estimators are particularly poor given how sensitive the
hold-out fits are to overfitting.
}
\label{fig:freq_pred_perf_overfit}
\end{figure*}

The potential pathologies of these approximations manifest even
in the simple misfit and overfit examples introduced above.
I generate an ensemble of data from the latent data generating process 
and compare the exact likelihood predictive performance score based 
on a reference Lebesgue measure, $\delta$, to each estimate, $\hat{\delta}$ 
(Figures \ref{fig:freq_pred_perf_misfit}, \ref{fig:freq_pred_perf_overfit}).  
The partitions for the hold-out estimators consisted of six data each, 
the minimum required for finite maximum likelihood estimates, while
the $K = 6$ jackknife partitions each consisting of $N - M = 10$ training 
data and $M = 2$ testing data.

In both cases the estimators are noisy with a substantial bias, with 
the hold-out estimator particularly sensitive to overfitting as expected.
Although these errors may partially cancel when comparing models,
any significant cancellation would be rather serendipitous.

\subsubsection{Comparing Prior Predictive Distributions}
\label{sec:comp_prior}

Box \cite{Box:1980, Box:1983} was a strong proponent of the
predictive validation of Bayesian methods, in particular the use
of the prior predictive distribution.

One benefit of the prior predictive distribution is that, because it
doesn't depend on the measurement, the delta estimator is
unbiased.  In fact, the delta estimate
\begin{equation*}
\delta_{D} \! \left( \pi \mid\mid \pi_{Y}^{\mathrm{prior}} \right) \approx
- \log \frac{ \mathrm{d} \pi_{Y}^{\mathrm{prior}} }{ \mathrm{d} \lambda } 
\! \left( y \right),
\end{equation*}
is exactly the logarithm of the marginal likelihood, or evidence, used in
Bayesian model comparison \cite{MacKay:2003, BernardoEtAl:2009},
and the difference of estimates between two models is exactly the 
log-odds ratio.  Consequently classical Bayesian model comparison 
also has an interpretation in terms of predictive performance.

That said, the utility of this relative predictive performance score
is limited both by the large variance of the estimator and a
potential overfitting bias if the prior is modified during inference.
In models where the prior is strongly constrained by previous
measurements or theoretical conditions this bias may be less
of an issue, but care should always be taken.  

\begin{figure*}
\centering
\subfigure[]{\includegraphics[width=2.9in]{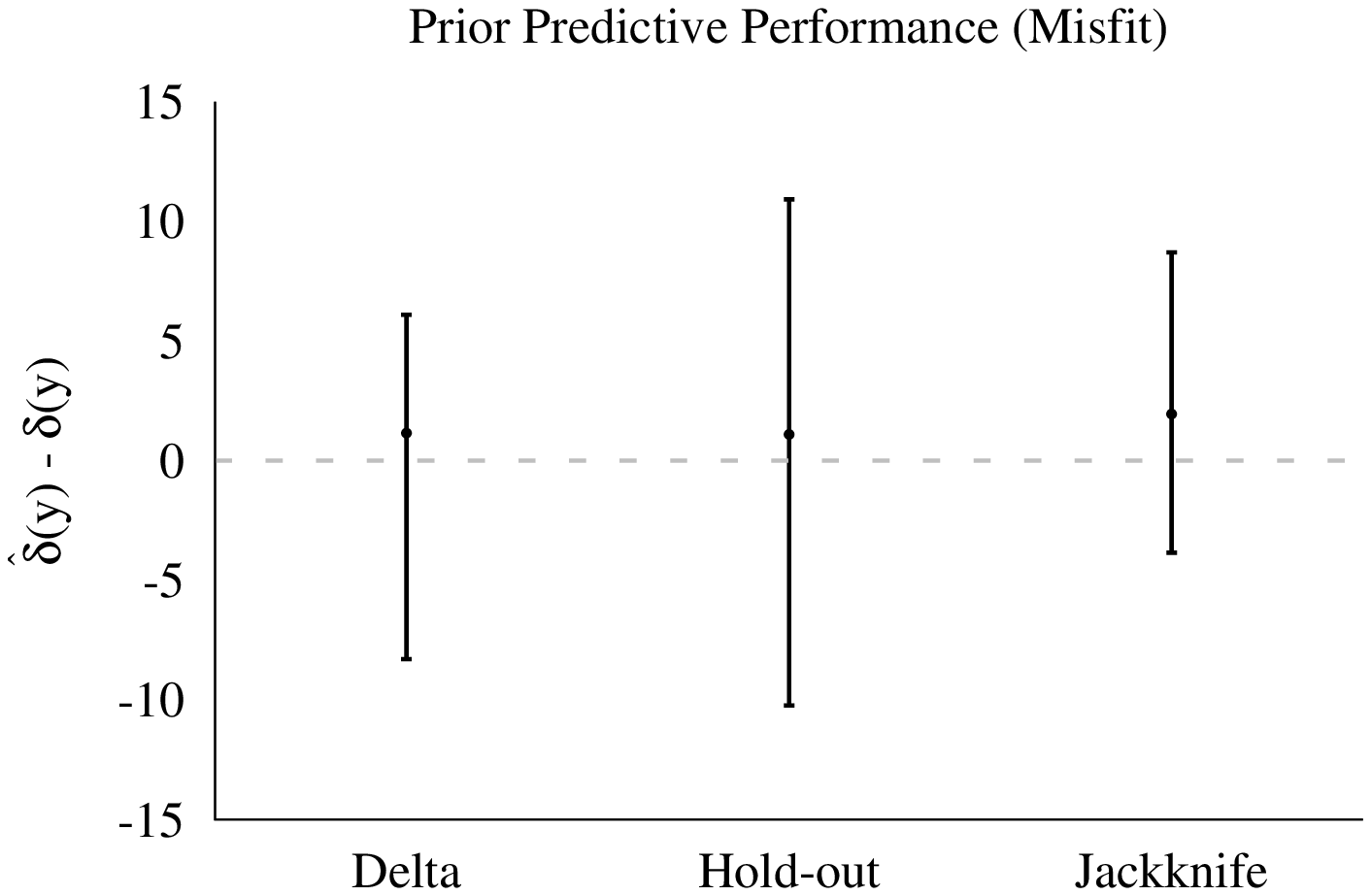}}
\subfigure[]{\includegraphics[width=2.9in]{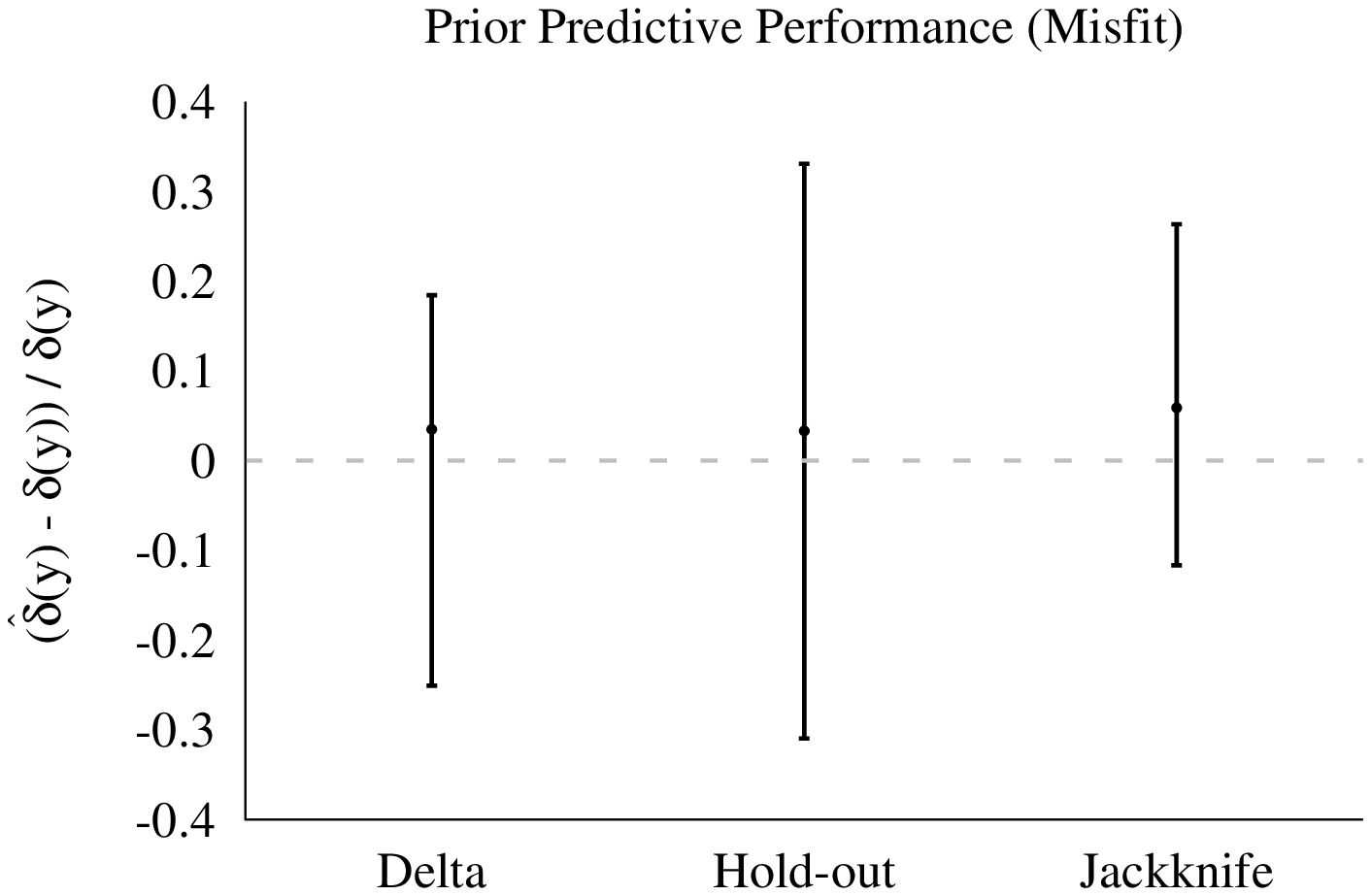}}
\caption{Approximations of prior predictive-based relative predictive 
performance scores for the misfit are not terrible, as 
demonstrated by the $20\%$, $50\%$, and $80\%$ quantiles of 
the estimator error over an ensemble of measurements from the 
latent data generating process, but nowhere near precise enough
to compare models with similar predictive performance.
}
\label{fig:prior_misfit_perf}
\end{figure*}

\begin{figure*}
\centering
\subfigure[]{\includegraphics[width=2.9in]{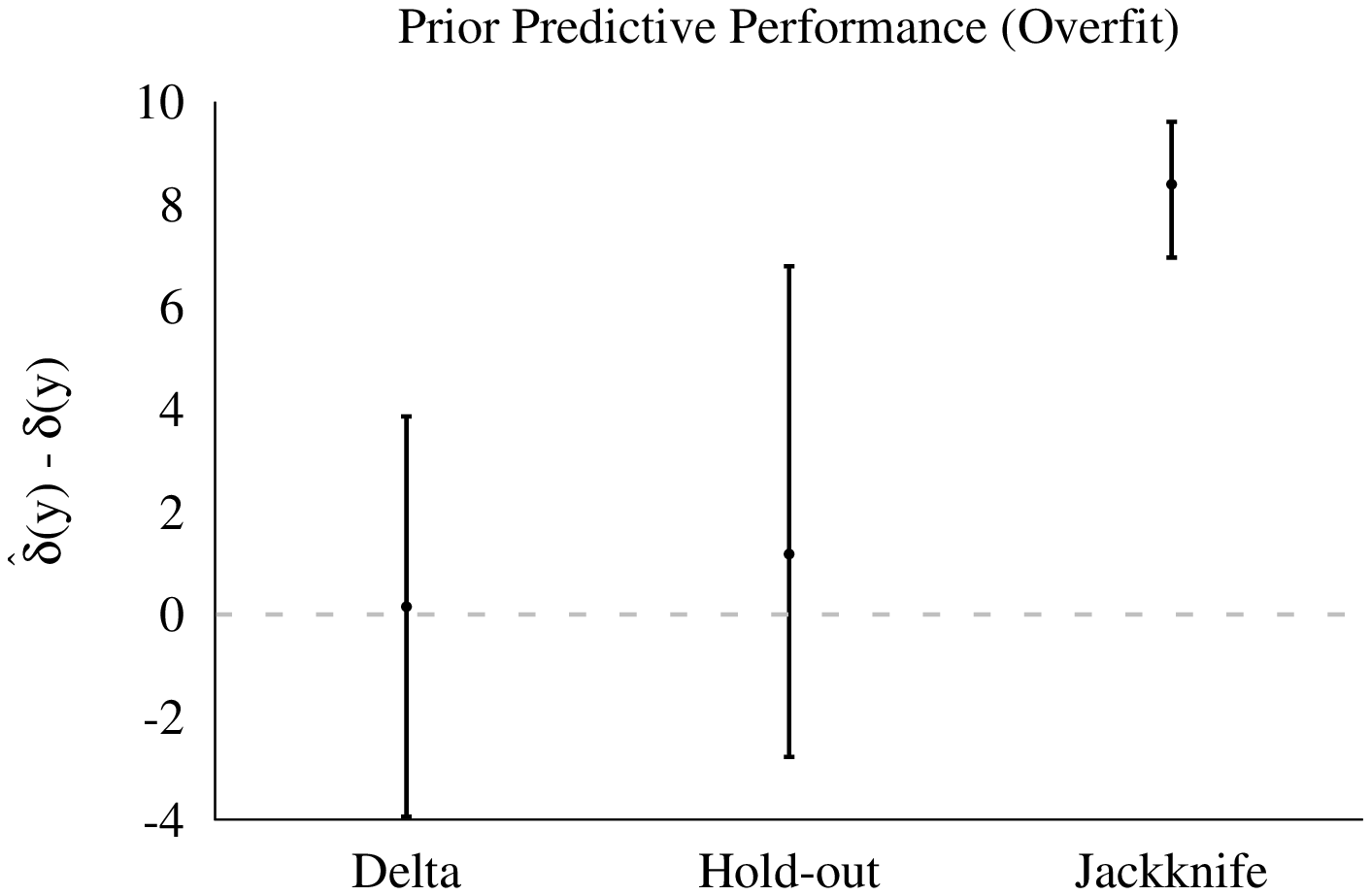}}
\subfigure[]{\includegraphics[width=2.9in]{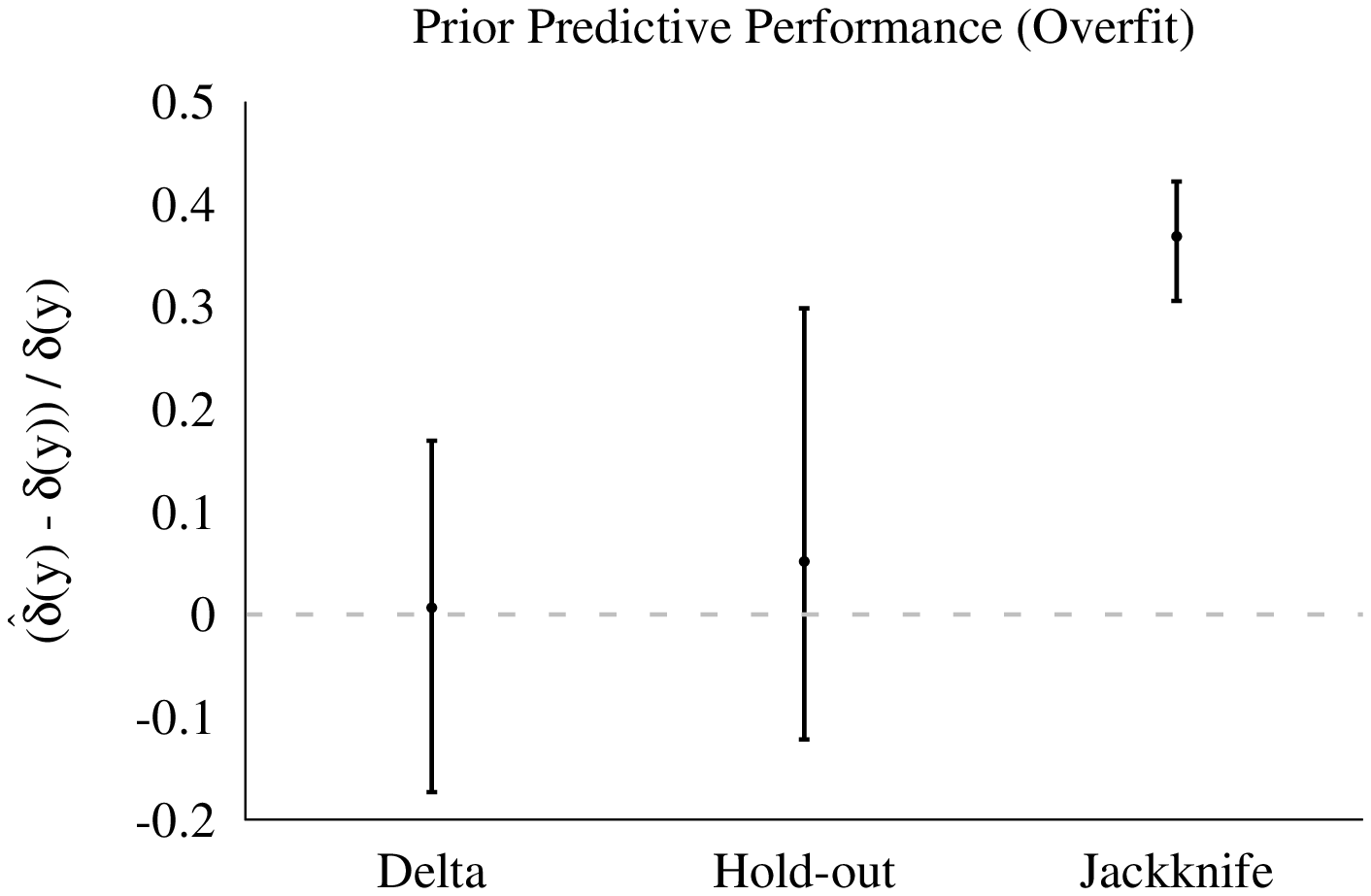}}
\caption{The bootstrap estimator of prior predictive-based relative 
predictive performance scores is particularly sensitive to overfitting,
as seen in the $20\%$, $50\%$, and $80\%$ quantiles of the estimator 
error over an ensemble of measurements from the latent data generating 
process.
}
\label{fig:prior_overfit_perf}
\end{figure*}

As in Section \ref{sec:comp_like}, the simple misfit and overfit
examples demonstrate the limitations of each estimator
(Figures \ref{fig:prior_misfit_perf}, \ref{fig:prior_overfit_perf}).

\subsubsection{Comparing Posterior Predictive Distributions}

Alternatively, we can construct relative predictive performance 
scores in the Bayesian paradigm by using the posterior predictive 
distribution.

\begin{figure*}
\centering
\subfigure[]{\includegraphics[width=2.9in]{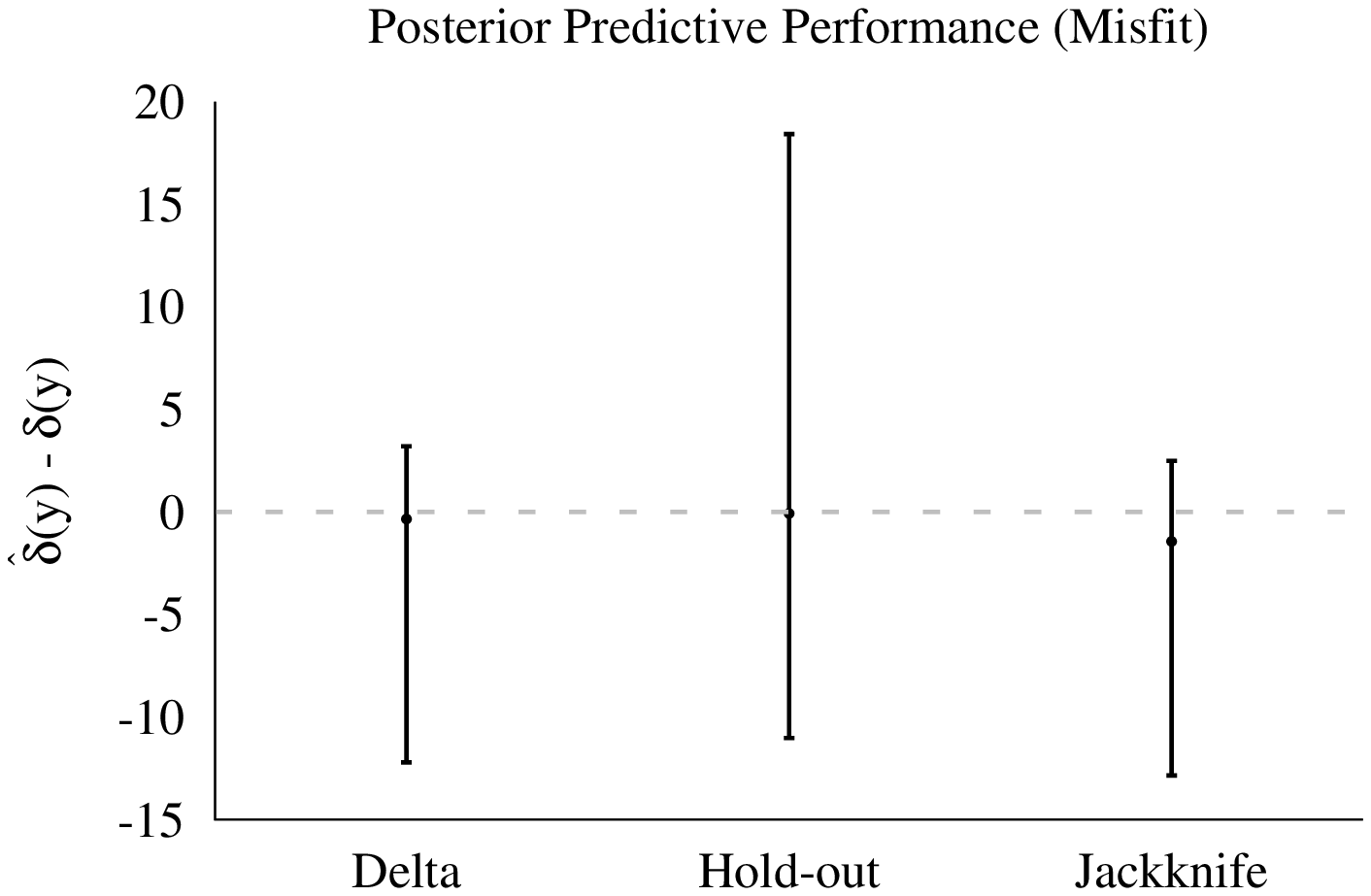}}
\subfigure[]{\includegraphics[width=2.9in]{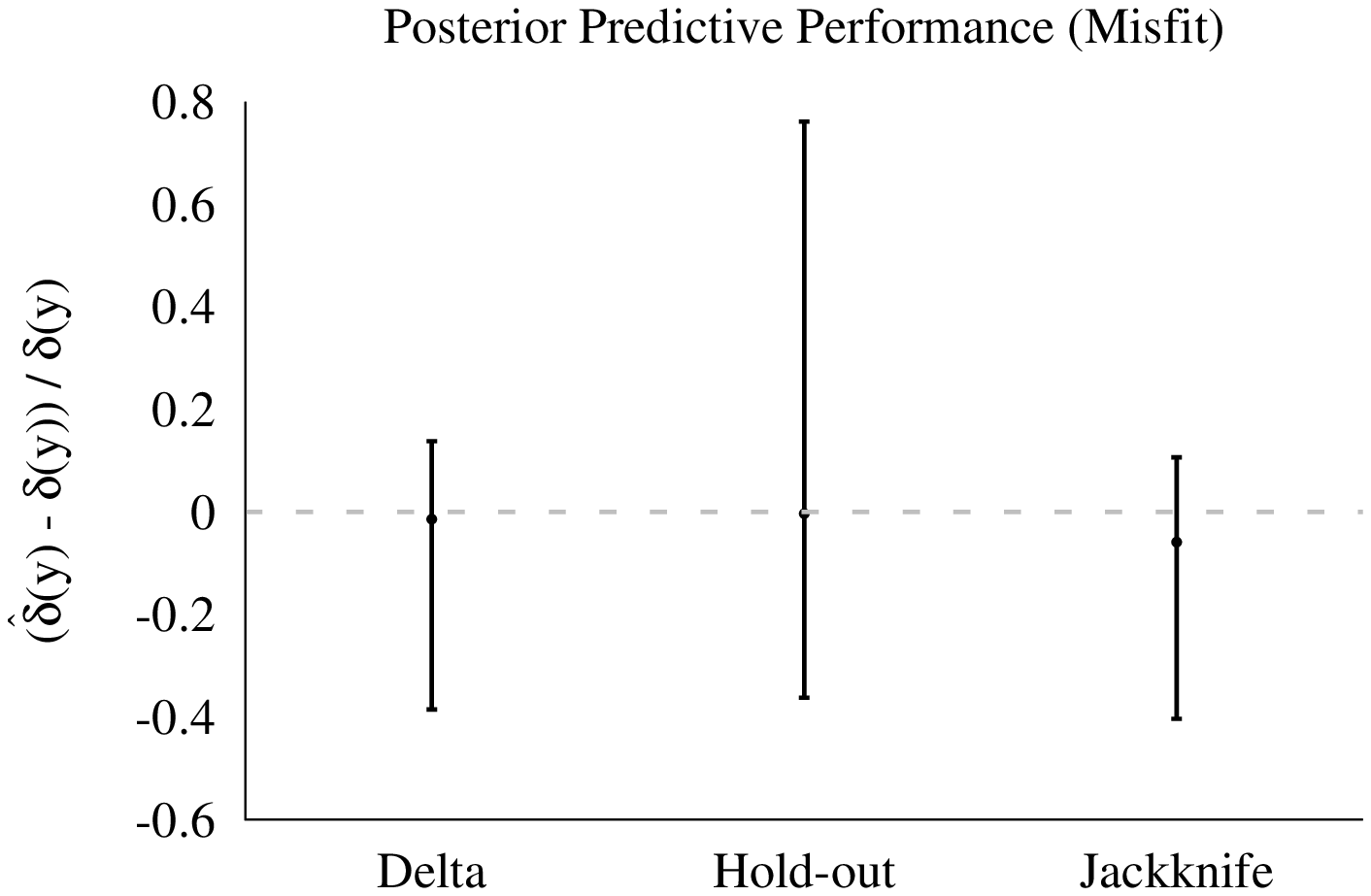}}
\caption{Approximations of posterior predictive-based relative predictive 
performance scores on the misfit example perform similarly to the 
approximations from other predictive distributions, once again 
demonstrated by the $20\%$, $50\%$, and $80\%$ quantiles of the 
estimator error over an ensemble of measurements from the latent 
data generating process.
}
\label{fig:post_misfit_perf}
\end{figure*}

\begin{figure*}
\centering
\subfigure[]{\includegraphics[width=2.9in]{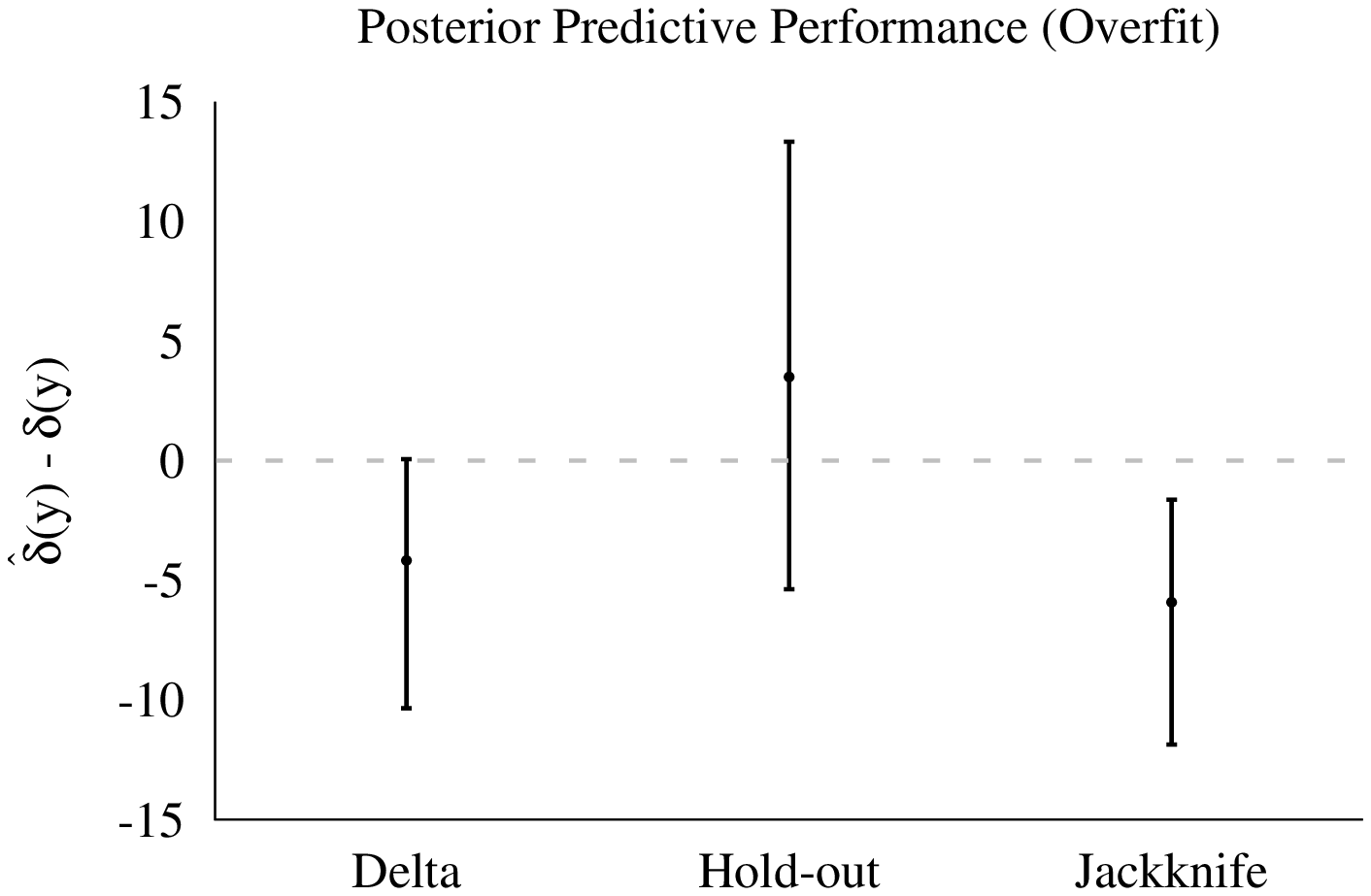}}
\subfigure[]{\includegraphics[width=2.9in]{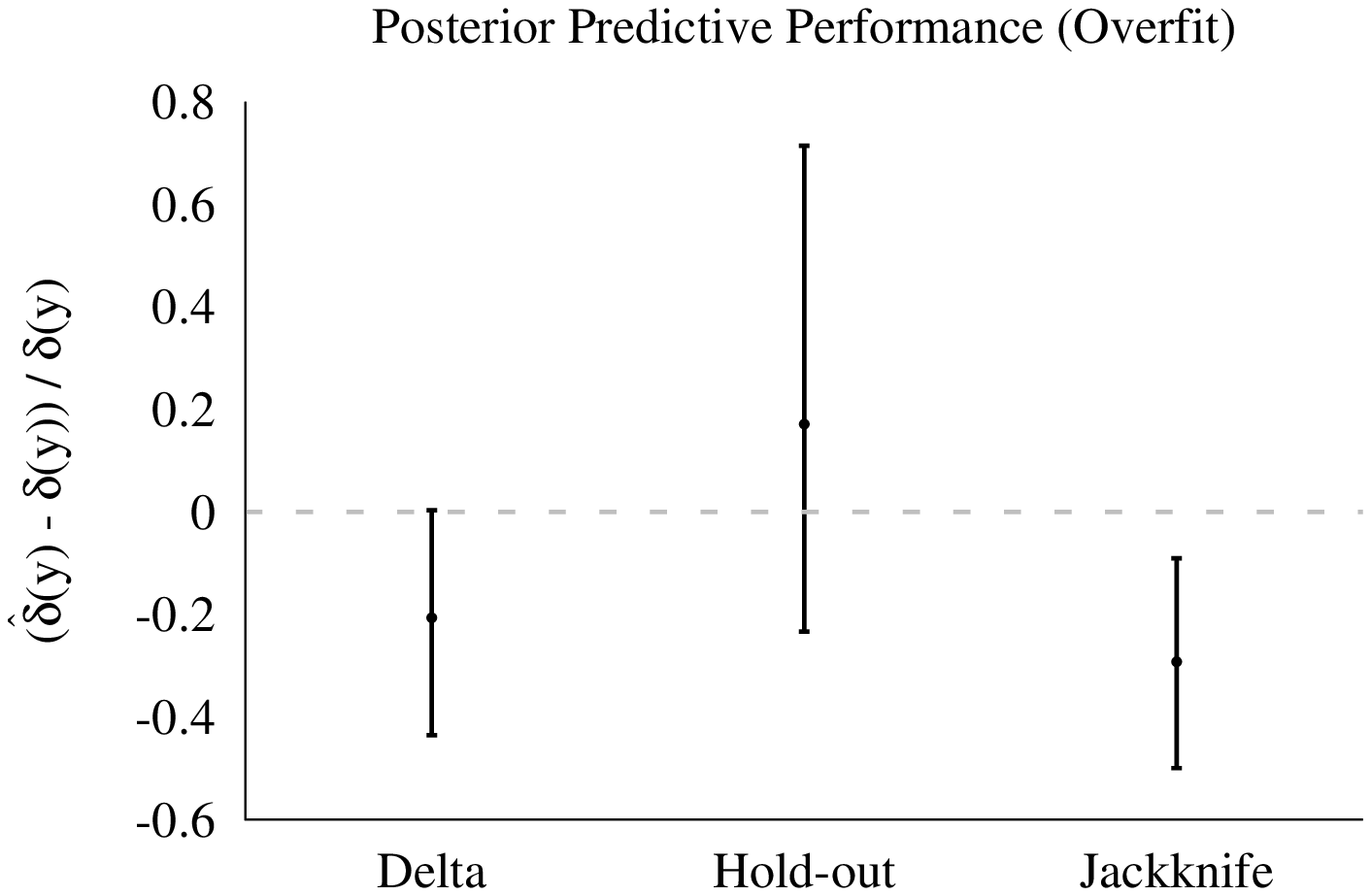}}
\caption{Approximations of posterior predictive-based relative predictive 
performance scores on the overfit example perform similarly to the 
approximations from other predictive distributions, once again 
demonstrated by the $20\%$, $50\%$, and $80\%$ quantiles of the 
estimator error over an ensemble of measurements from the latent 
data generating process.
}
\label{fig:post_overfit_perf}
\end{figure*}

Similar to the use of likelihoods, relative predictive performance scores 
constructed from reuse estimators provide motivation for many visual
diagnostics such as Bayesian residual analyses \cite{Gelman:2003}
and, in particular, posterior-predictive checks \cite{Rubin:1981, Rubin:1984}.
Likewise, the use of hold-out and jackknife estimators yields posterior 
predictive hold-out and cross validation, \cite{GelfandEtAl:1992, Gelfand:1996}, 
which continues to grow in popularity in the machine learning and statistics 
literature.  Consideration of the example of Section \ref{sec:comp_like}
and \ref{sec:comp_prior} emphasizes the continued need to maintain 
vigilance in these applications (Figures \ref{fig:post_misfit_perf},
\ref{fig:post_overfit_perf}).

Posterior predictive cross validation also provides the basis for unifying 
many of the information criteria that have been developed in the Bayesian 
literature.  \cite{VehtariEtAl:2012, GelmanEtAl:2014b}, for example, show 
how the Widely Applicable Information Criterion \cite{Watanabe:2013},
\begin{equation*}
\mathrm{WAIC} \propto 
\sum_{n = 1}^{N} \log \mathbb{E}_{x} \left[ \pi_{Y | x} \! \left( y_{n} | x \right) \right]
- \sum_{n = 1}^{N} \mathrm{Var}_{x} \left[ \log \pi_{Y | x} \! \left( y_{n} | x \right) \right],
\end{equation*}
can be derived as an approximation of the posterior predictive relative 
predictive performance score.  Moreover, given a point estimate, $\hat{x}$, 
that singles out one element of the small world, the Widely Applicable Information 
Criterion reduces to the Deviance Information Criterion \cite{SpiegelhalterEtAl:2002}, 
\begin{equation*}
\mathrm{DIC} \propto
\sum_{n = 1}^{N} \log \pi_{Y | x} \! \left( y_{n} | \hat{x} \right)
- 2 \sum_{n = 1}^{N} \left(
\log \pi_{Y | x} \! \left( y_{n} | \hat{x} \right)
- \mathbb{E}_{x} \! \left[ \log \pi_{Y | x} \! \left( y_{n} | \hat{x} \right) \right] \right).
\end{equation*}
As with the reuse, hold-out, and jackknife estimators, the ultimate 
accuracy and precision of these information criteria relative to the true 
posterior predictive relative predictive performance score is paramount
in any partial application.

\section{Conclusion}

In this paper I have shown how the Kullback-Leibler divergence between an 
inferred predictive distribution and the latent data generating process defines 
a canonical but incomputable measure of relative predictive performance. 
Moreover, I have demonstrated how it can be simplified into relative predictive 
performance scores which quantify the same relative predictive performance 
while also being more amenable to practical approximations.  
Applying various approximation strategies to the relative predictive performance
derived from predictive distributions in frequentist and Bayesian inference 
yields many of the model comparison techniques ubiquitous in practice,
from predictive log loss cross validation to the Bayesian evidence and
Bayesian information criteria.

The main benefit in unifying all of these existing methods into a single 
foundational perspective is that it provides a common framework for 
understanding the limitations of these methods and how they can be used 
responsibly.  In particular, it emphasizes that these existing methods are 
all estimates, with uncertain variances and biases that make quantitive 
statements about relative predictive performance, and a hard selection 
of one model above all others under consideration, somewhat precarious.  

This difficulty in making quantitative statements has motivated new 
approaches to model comparison that do not fit each model in isolation 
but rather fit them as components of a single, comprehensive model 
\cite{KamaryEtAl:2014, PiironenEtAl:2015}.
Such an inclusive strategy offers unique computational benefits and
an intriguing new interpretation of the small world, and it promises to
be an exciting area of future research.

\section*{Acknowledgements}

The fundamental perspective in this paper attempts to unity the
work of many others, whose own contributions cannot be overlooked.
I am particularly thankful for long conversations with Simon Byrne,
Bob Carpenter, Andrew Gelman, Sarah Heaps, Christian Robert, 
and Aki Vehtari.  
This work was supported under EPSRC grant EP/J016934/1.

\bibliography{unified_predictive_inference}
\bibliographystyle{PhysRevStyle}

\end{document}